\documentclass[twocolumn]{aastex631}
\usepackage{amsmath}
\usepackage{epsf}
\usepackage{color}
\usepackage{color}
\usepackage{graphicx}
\usepackage{color}
\usepackage{enumitem}
\usepackage{wrapfig}
\usepackage{mathtools}
\usepackage[mathcal]{eucal}
 \let\mathscr\relax
\usepackage[scr]{rsfso}

\usepackage{natbib}

\shorttitle{CEERS Survey Overview}
\shortauthors{Finkelstein et al.}
\newcommand{\sol}{$_{\odot}$}

\newcommand{\hst}{{\it HST}}

\def\arcs{\hbox{$^{\prime\prime}$}}

\newcommand{\CIV}{\textrm{C}\,\textsc{iv}}
\newcommand{\CIII}{\textrm{C}\,\textsc{iii}\textrm{]}}
\newcommand{\OII}{\textrm{[O}\,\textsc{ii}\textrm{]}}

\newcommand{\NeV}{\textrm{[Ne}\,\textsc{v}\textrm{]}}
\newcommand{\OIII}{\textrm{[O}\,\textsc{iii}\textrm{]}}

\newcommand{\HeII}{\textrm{He}\,\textsc{ii}}
\newcommand{\HII}{\textrm{H}\,\textsc{ii}}
\newcommand{\ergs}{\ifmmode \textrm{erg\ s}^{-1} \else erg s$^{-1}$\fi}
\newcommand{\Msun}{\ifmmode \textrm{M}_{\odot} \else $M_{\odot}$\fi}
\newcommand{\Lsun}{\ifmmode \textrm{L}_{\odot} \else $L_{\odot}$\fi}

\newcommand{\JWST}{\textit{JWST}}

\definecolor{aggiemaroon}{HTML}{500000}

\begin{document}
\title{The Cosmic Evolution Early Release Science Survey (CEERS)}

% TAH moving affilliations to end of document
% to clean up first page (so abstract is actually on first page)
\suppressAffiliations

\author[0000-0001-8519-1130]{Steven L. Finkelstein}
\affiliation{Department of Astronomy, The University of Texas at Austin, Austin, TX, USA}
\email{stevenf@astro.as.utexas.edu}

\author[0000-0002-9921-9218]{Micaela B. Bagley}
\affiliation{Department of Astronomy, The University of Texas at Austin, Austin, TX, USA}
\affiliation{Astrophysics Science Division, NASA Goddard Space Flight Center, 8800 Greenbelt Rd, Greenbelt, MD 20771, USA}

%%%%%%%%%%%%%%%%%%%%%%%%%%%%%%%%%%%%%%%%%%%%%%%%%%%%%%%%%%%%%%%%%%%%%%%%%%%%%%
%%% Program-level architects, key proposal authors %%%
\author[0000-0002-7959-8783]{Pablo Arrabal Haro}
\affiliation{NSF's National Optical-Infrared Astronomy Research Laboratory, 950 N. Cherry Ave., Tucson, AZ 85719, USA}

\author[0000-0001-5414-5131]{Mark Dickinson}
\affiliation{NSF's National Optical-Infrared Astronomy Research Laboratory, 950 N. Cherry Ave., Tucson, AZ 85719, USA}

\author[0000-0001-7113-2738]{Henry C. Ferguson}
\affiliation{Space Telescope Science Institute, Baltimore, MD, USA}

\author[0000-0001-9187-3605]{Jeyhan S. Kartaltepe}
\affiliation{Laboratory for Multiwavelength Astrophysics, School of Physics and Astronomy, Rochester Institute of Technology, 84 Lomb Memorial Drive, Rochester, NY 14623, USA}

\author[0000-0002-8360-3880]{Dale D. Kocevski}
\affiliation{Department of Physics and Astronomy, Colby College, Waterville, ME 04901, USA}

\author[0000-0002-6610-2048]{Anton M. Koekemoer}
\affiliation{Space Telescope Science Institute, 3700 San Martin Drive, Baltimore, MD 21218, USA}

\author[0000-0003-3130-5643]{Jennifer M. Lotz}
\affiliation{Gemini Observatory/NSF's National Optical-Infrared Astronomy Research Laboratory, 950 N. Cherry Ave., Tucson, AZ 85719, USA}
\affiliation{Space Telescope Science Institute, 3700 San Martin Drive, Baltimore, MD 21218, USA}

\author[0000-0001-7503-8482]{Casey Papovich}
\affiliation{George P. and Cynthia Woods Mitchell Institute for Fundamental Physics and Astronomy, Department of Physics and Astronomy, Texas A\&M University, College Station, TX, USA}

\author[0000-0003-4528-5639]{Pablo G. P\'erez-Gonz\'alez}
\affiliation{Centro de Astrobiolog\'{\i}a (CAB/CSIC-INTA), Ctra. de Ajalvir km 4, Torrej\'on de Ardoz, E-28850, Madrid, Spain}

\author[0000-0003-3382-5941]{Nor Pirzkal}
\affiliation{ESA/AURA Space Telescope Science Institute}

\author[0000-0002-6748-6821]{Rachel S.~Somerville}
\affiliation{Center for Computational Astrophysics, Flatiron Institute, 162 5th Avenue, New York, NY 10010, USA}

\author[0000-0002-1410-0470]{Jonathan R. Trump}
\affiliation{Department of Physics, 196A Auditorium Road, Unit 3046, University of Connecticut, Storrs, CT 06269, USA}

\author[0000-0001-8835-7722]{Guang Yang}
\affiliation{Nanjing Institute of Astronomical Optics and Technology, Nanjing 210042, China}

\author[0000-0003-3466-035X]{{L. Y. Aaron} {Yung}}
\affiliation{Space Telescope Science Institute, 3700 San Martin Drive, Baltimore, MD 21218, USA}

%%%%%%%%%%%%%%%%%%%%%%%%%%%%%%%%%%%%%%%%%%%%%%%%%%%%%%%%%%%%%%%%%%%%%%%%%%%%%%
%%Key project architects not included above and other significant contributors to the paper content%%
\author[0000-0003-3820-2823]{Adriano Fontana}
\affiliation{INAF - Osservatorio Astronomico di Roma, via di Frascati 33, 00078 Monte Porzio Catone, Italy}

\author[0000-0002-5688-0663]{Andrea Grazian}
\affiliation{INAF--Osservatorio Astronomico di Padova, Vicolo dell'Osservatorio 5, I-35122, Padova, Italy}

\author[0000-0001-9440-8872]{Norman A. Grogin}
\affiliation{Space Telescope Science Institute, 3700 San Martin Drive, Baltimore, MD 21218, USA}

\author[0000-0001-8152-3943]{Lisa J. Kewley}
\affiliation{Harvard-Smithsonian Center for Astrophysics, 60 Garden Street, Cambridge, MA 02138, USA}

\author[0000-0002-5537-8110]{Allison Kirkpatrick}
\affiliation{Department of Physics and Astronomy, University of Kansas, Lawrence, KS 66045, USA}

\author[0000-0003-2366-8858]{Rebecca L. Larson}
\affiliation{Laboratory for Multiwavelength Astrophysics, School of Physics and Astronomy, Rochester Institute of Technology, 84 Lomb Memorial Drive, Rochester, NY 14623, USA}

\author[0000-0001-8940-6768]{Laura Pentericci}
\affiliation{INAF - Osservatorio Astronomico di Roma, via di Frascati 33, 00078 Monte Porzio Catone, Italy}

\author[0000-0002-5269-6527]{Swara Ravindranath}
\affiliation{Space Telescope Science Institute, 3700 San Martin Drive, Baltimore, MD 21218, USA}

\author[0000-0003-3903-6935]{Stephen M.~Wilkins} %
\affiliation{Astronomy Centre, University of Sussex, Falmer, Brighton BN1 9QH, UK}
\affiliation{Institute of Space Sciences and Astronomy, University of Malta, Msida MSD 2080, Malta}

%%%%%%%%%%%%%%%%%%%%%%%%%%%%%%%%%%%%%%%%%%%%%%%%%%%%%%%%%%%%%%%%%%%%%%%%%%%%%%
%%%%%%%%%%%%%%%%%%%%%%   DO NOT EDIT ABOVE  %%%%%%%%%%%%%%%%%%%%%%
%%%%%%%%%%%%%%%%%%%%%%%%%%%%%%%%%%%%%%%%%%%%%%%%%%%%%%%%%%%%%%%%%%%%%%%%%%%%%%

%%%%%%%%%%%%%%%%%%%%%%%%%%%%%%%%%%%%%%%%%%%%%%%%%%%%%%%%%%%%%%%%%%%%%%%%%%%%%%
%%----TEAM MEMBERS ENTER AUTHOR BLOCK INFO BELOW------%%

\author[0000-0001-9328-3991]{Omar Almaini}
\affiliation{School of Physics and Astronomy, University of Nottingham, University Park, Nottingham NG7 2RD, UK}

\author[0000-0001-5758-1000]{Ricardo O. Amor\'{i}n}
\affiliation{Instituto de Astrof\'{i}sica de Andaluc\'{i}a (CSIC), Apartado 3004, 18080 Granada, Spain}

\author[0000-0001-6813-875X]{Guillermo Barro}
\affiliation{University of the Pacific, Stockton, CA 90340 USA}

\author[0000-0003-0883-2226]{Rachana Bhatawdekar}
\affiliation{European Space Agency (ESA), European Space Astronomy Centre (ESAC), Camino Bajo del Castillo s/n, 28692 Villanueva de la Cañada, Madrid, Spain}

\author[0000-0003-0492-4924]{Laura Bisigello}
\affiliation{INAF, Istituto di Radioastronomia, Via Piero Gobetti 101, 40129 Bologna, Italy}
\affiliation{Dipartimento di Fisica e Astronomia "G.Galilei", Universit\'a di Padova, Via Marzolo 8, I-35131 Padova, Italy}

\author[0000-0001-5384-3616]{Madisyn Brooks}
\affiliation{Department of Physics, 196A Auditorium Road, Unit 3046, University of Connecticut, Storrs, CT 06269, USA}

\author[0000-0002-2861-9812]{Fernando Buitrago}
\affiliation{Departamento de F\'{i}sica Te\'{o}rica, At\'{o}mica y \'{O}ptica, Universidad de Valladolid, 47011 Valladolid, Spain}
\affiliation{Instituto de Astrof\'{i}sica e Ci\^{e}ncias do Espa\c{c}o, Universidade de Lisboa, OAL, Tapada da Ajuda, PT1349-018 Lisbon, Portugal}

\author[0000-0003-2536-1614]{Antonello Calabr\`o}
\affiliation{INAF Osservatorio Astronomico di Roma, Via Frascati 33, 00078 Monte Porzio Catone, Italy}

\author[0000-0001-9875-8263]{Marco Castellano}
\affiliation{INAF - Osservatorio Astronomico di Roma, via di Frascati 33, 00078 Monte Porzio Catone, Italy}

\author[0000-0001-8551-071X]{Yingjie Cheng}
\affiliation{University of Massachusetts Amherst \\
710 North Pleasant Street, Amherst, MA 01003-9305, USA}

\author[0000-0001-7151-009X]{Nikko J. Cleri}
\affiliation{Department of Astronomy and Astrophysics, The Pennsylvania State University, University Park, PA 16802, USA}
\affiliation{Institute for Computational and Data Sciences, The Pennsylvania State University, University Park, PA 16802, USA}
\affiliation{Institute for Gravitation and the Cosmos, The Pennsylvania State University, University Park, PA 16802, USA}

\author[0000-0002-6348-1900]{Justin W. Cole}
\altaffiliation{NASA FINESST Investigator}
\affiliation{Department of Physics and Astronomy, Texas A\&M
  University, College Station, TX, 77843-4242 USA}
\affiliation{George P.\ and Cynthia Woods Mitchell Institute for
  Fundamental Physics and Astronomy, Texas A\&M University, College
  Station, TX, 77843-4242 USA}

\author[0000-0003-1371-6019]{M. C. Cooper}
\affiliation{Department of Physics \& Astronomy, University of California, Irvine, 4129 Reines Hall, Irvine, CA 92697, USA}

\author[0000-0003-3881-1397]{Olivia R. Cooper}\altaffiliation{NSF Graduate Research Fellow}
\affiliation{Department of Astronomy, The University of Texas at Austin, Austin, TX, USA}

\author[0000-0001-6820-0015]{Luca Costantin}
\affiliation{Centro de Astrobiolog\'{\i}a (CAB), CSIC-INTA, Ctra. de Ajalvir km 4, Torrej\'on de Ardoz, E-28850, Madrid, Spain}

\author[0000-0002-1803-794X]{Isa G. Cox}
\affiliation{Laboratory for Multiwavelength Astrophysics, School of Physics and Astronomy, Rochester Institute of Technology, 84 Lomb Memorial Drive, Rochester, NY
14623, USA}

\author[0000-0002-5009-512X]{Darren Croton}
\affiliation{Centre for Astrophysics \& Supercomputing, Swinburne University of Technology, Hawthorn, VIC 3122, Australia}
\affiliation{ARC Centre of Excellence for All Sky Astrophysics in 3 Dimensions (ASTRO 3D)}
\affiliation{ARC Centre of Excellence for Dark Matter Particle Physics (CDM)}

\author[0000-0002-3331-9590]{Emanuele Daddi}
\affiliation{Universit{\'e} Paris-Saclay, Universit{\'e} Paris Cit{\'e}, CEA, CNRS, AIM, 91191, Gif-sur-Yvette, France}

\author[0000-0001-8047-8351]{Kelcey Davis}
\altaffiliation{NSF Graduate Research Fellow}
\affiliation{Department of Physics, 196A Auditorium Road, Unit 3046, University of Connecticut, Storrs, CT 06269, USA}

\author[0000-0003-4174-0374]{Avishai Dekel}
\affiliation{Racah Institute of Physics, The Hebrew University of Jerusalem, Jerusalem 91904, Israel}

\author[0000-0002-7631-647X]{David Elbaz}
\affiliation{Universit\'e Paris-Saclay, Universit\'e Paris Cit\'e, CEA, CNRS, AIM, 91191, Gif-sur-Yvette, France}

\author[0000-0003-0531-5450]{Vital Fern\'andez}
\affiliation{Michigan Institute for Data Science, Unversity of Michigan, 500 Church Street, Ann Arbor, MI 48109, USA}

\author[0000-0001-7201-5066]{Seiji Fujimoto}
\altaffiliation{Hubble Fellow}
\affiliation{Department of Astronomy, The University of Texas at Austin, Austin, TX, USA}

\author[0000-0003-3248-5666]{Giovanni Gandolfi}
\affiliation{Dipartimento di Fisica e Astronomia "G.Galilei", Universit\'a di Padova, Via Marzolo 8, I-35131 Padova, Italy}
\affiliation{INAF--Osservatorio Astronomico di Padova, Vicolo dell'Osservatorio 5, I-35122, Padova, Italy}

\author[0000-0003-2098-9568]{Jonathan P. Gardner}
\affiliation{Sciences and Exploration Directorate, NASA Goddard Space Flight Center, 8800 Greenbelt Rd, Greenbelt, MD 20771, USA}

\author[0000-0003-1530-8713]{Eric Gawiser}
\affiliation{Department of Physics and Astronomy, Rutgers, the State University of New Jersey, Piscataway, NJ 08854, USA}

\author[0000-0002-7831-8751]{Mauro Giavalisco}
\affiliation{University of Massachusetts Amherst \\
710 North Pleasant Street, Amherst, MA 01003-9305, USA}

\author[0000-0002-4085-9165]{Carlos G{\'o}mez-Guijarro}
\affiliation{Universit{\'e} Paris-Saclay, Universit{\'e} Paris Cit{\'e}, CEA, CNRS, AIM, 91191, Gif-sur-Yvette, France}

\author[0000-0002-4162-6523]{Yuchen Guo}
\affiliation{Department of Astronomy, The University of Texas at Austin, Austin, TX, USA}

\author[0000-0003-4242-8606]{Ansh R. Gupta}\altaffiliation{NSF Graduate Research Fellow}
\affiliation{Department of Astronomy, The University of Texas at Austin, Austin, TX, USA}

\author[0000-0001-6145-5090]{Nimish P. Hathi}
\affiliation{Space Telescope Science Institute, 3700 San Martin Drive, Baltimore, MD 21218, USA}

\author[0000-0003-0129-2079]{Santosh Harish}
\affiliation{Laboratory for Multiwavelength Astrophysics, School of Physics and Astronomy, Rochester Institute of Technology, 84 Lomb Memorial Drive, Rochester, NY 14623, USA}

\author[0000-0002-9466-2763]{Aur{\'e}lien Henry}
\affiliation{Department of Physics, University of California, Merced, 5200 Lake Road, Merced, CA 92543, USA}

\author[0000-0002-3301-3321]{Michaela Hirschmann}
\affiliation{Institute of Physics, Laboratory of Galaxy Evolution, Ecole Polytechnique Federale de Lausanne (EPFL), Observatoire de Sauverny, 1290 Versoix, Switzerland}

\author[0000-0003-3424-3230]{Weida Hu}
\affiliation{Department of Physics and Astronomy, Texas A\&M University, College Station, TX 77843-4242, USA}
\affiliation{George P. and Cynthia Woods Mitchell Institute for Fundamental Physics and Astronomy, Texas A\&M University, College Station, TX 77843-4242, USA}

\author[0000-0001-6251-4988]{Taylor A. Hutchison}
\altaffiliation{NASA Postdoctoral Fellow}
\affiliation{Astrophysics Science Division, NASA Goddard Space Flight Center, 8800 Greenbelt Rd, Greenbelt, MD 20771, USA}

\author[0000-0001-9298-3523]{Kartheik G. Iyer}
\altaffiliation{Hubble Fellow}
\affiliation{Columbia Astrophysics Laboratory, Columbia University, 550 West 120th Street, New York, NY 10027, USA}

\author[0000-0002-6790-5125]{Anne E. Jaskot}
\affiliation{Williams College, 33 Lab Campus Drive, Williamstown, MA 01267, USA}

\author[0000-0001-8738-6011]{Saurabh~W.~Jha}
\affiliation{Department of Physics and Astronomy, Rutgers, the State University of New Jersey, Piscataway, NJ 08854, USA}

\author[0000-0003-1187-4240]{Intae Jung}
\affiliation{Space Telescope Science Institute, 3700 San Martin Drive Baltimore, MD 21218, United States}

\author[0000-0002-5588-9156]{Vasily Kokorev}
\affiliation{Department of Astronomy, The University of Texas at Austin, Austin, TX, USA}

\author[0000-0002-8816-5146]{Peter Kurczynski}
\affiliation{NASA Goddard Space Flight Center, Greenbelt MD 20771, USA}

\author[0000-0002-9393-6507]{Gene C. K. Leung}
\affiliation{MIT Kavli Institute for Astrophysics and Space Research, 77 Massachusetts Ave., Cambridge, MA 02139, USA}

\author[0000-0003-1354-4296]{Mario Llerena}
\affiliation{INAF - Osservatorio Astronomico di Roma, via di Frascati 33, 00078 Monte Porzio Catone, Italy}

\author[0000-0002-7530-8857]{Arianna S. Long}
\affiliation{Department of Astronomy, The University of Washington, Seattle, WA 98195, USA}

\author[0000-0003-1581-7825]{Ray A. Lucas}
\affiliation{Space Telescope Science Institute, 3700 San Martin Drive, Baltimore, MD 21218, USA}

\author[0000-0001-5988-2202]{Shiying Lu}
\affiliation{School of Astronomy and Space Science, Nanjing University, Nanjing, 210093,China}
\affiliation{Key Laboratory of Modern Astronomy and Astrophysics (Nanjing University), Ministry of Education, Nanjing, 210093, China}

\author[0000-0001-8688-2443]{Elizabeth J.\ McGrath}
\affiliation{Department of Physics and Astronomy, Colby College, Waterville, ME 04901, USA}

\author[0009-0000-1182-6420]{Daniel H.\ McIntosh}
\affiliation{Division of Energy, Matter and Systems, School of Science and Engineering, University of Missouri, Kansas City, MO 64110, USA}

\author[0000-0001-6870-8900]{Emiliano Merlin}
\affiliation{INAF Osservatorio Astronomico di Roma, Via Frascati 33, 00078 Monteporzio Catone, Rome, Italy}

\author[0000-0003-4965-0402]{Alexa M.\ Morales}\altaffiliation{NSF Graduate Research Fellow}
\affil{Department of Astronomy, The University of Texas at Austin, 2515 Speedway, Austin, TX, 78712, USA}

\author[0000-0002-8951-4408]{Lorenzo Napolitano}
\affiliation{INAF – Osservatorio Astronomico di Roma, via Frascati 33, 00078, Monteporzio Catone, Italy}
\affiliation{Dipartimento di Fisica, Università di Roma Sapienza, Città Universitaria di Roma - Sapienza, Piazzale Aldo Moro, 2, 00185, Roma, Italy}

\author[0000-0001-9879-7780]{Fabio Pacucci}
\affiliation{Center for Astrophysics $\vert$ Harvard \& Smithsonian, 60 Garden St, Cambridge, MA 02138, USA}
\affiliation{Black Hole Initiative, Harvard University, 20 Garden St, Cambridge, MA 02138, USA}

\author[0000-0002-2499-9205]{Viraj Pandya}
\altaffiliation{Hubble Fellow}
\affiliation{Columbia Astrophysics Laboratory, Columbia University, 550 West 120th Street, New York, NY 10027, USA}

\author[0000-0002-9946-4731]{Marc Rafelski}
\affiliation{Space Telescope Science Institute, 3700 San Martin Drive, Baltimore, MD 21218, USA}
\affiliation{Department of Physics and Astronomy, Johns Hopkins University, Baltimore, MD 21218, USA}

\author[0000-0002-9415-2296]{Giulia Rodighiero}
\affiliation{Dipartimento di Fisica e Astronomia "G.Galilei", Universit\'a di Padova, Via Marzolo 8, I-35131 Padova, Italy}
\affiliation{INAF--Osservatorio Astronomico di Padova, Vicolo dell'Osservatorio 5, I-35122, Padova, Italy}

\author[0000-0002-8018-3219]{Caitlin Rose}
\affil{Laboratory for Multiwavelength Astrophysics, School of Physics and Astronomy, Rochester Institute of Technology, 84 Lomb Memorial Drive, Rochester, NY 14623, USA}

\author[0000-0002-9334-8705]{Paola Santini}
\affiliation{INAF - Osservatorio Astronomico di Roma, via di Frascati 33, 00078 Monte Porzio Catone, Italy}

\author[0000-0001-7755-4755]{Lise-Marie Seillé}
\affiliation{Aix Marseille Univ, CNRS, CNES, LAM, Marseille, France}

\author[0000-0002-6386-7299]{Raymond C.\ Simons}
\affiliation{Department of Engineering and Physics, Providence College, 1 Cunningham Sq, Providence, RI 02918 USA}

\author[0000-0001-9495-7759]{Lu Shen}
\affiliation{Department of Physics and Astronomy, Texas A\&M University, College Station, TX, 77843-4242 USA}
\affiliation{George P.\ and Cynthia Woods Mitchell Institute for
 Fundamental Physics and Astronomy, Texas A\&M University, College Station, TX, 77843-4242 USA}

\author[0000-0002-4772-7878]{Amber N. Straughn}
\affiliation{Astrophysics Science Division, NASA Goddard Space Flight Center, 8800 Greenbelt Rd, Greenbelt, MD 20771, USA}
 
\author[0000-0002-8224-4505]{Sandro Tacchella}
\affiliation{Kavli Institute for Cosmology, University of Cambridge, Madingley Road, Cambridge, CB3 0HA, UK}
\affiliation{Cavendish Laboratory - Astrophysics Group, University of Cambridge, 19 JJ Thomson Avenue, Cambridge, CB3 0HE, UK}

\author[0000-0002-8163-0172]{Brittany N. Vanderhoof}
\affil{Laboratory for Multiwavelength Astrophysics, School of Physics and Astronomy, Rochester Institute of Technology, 84 Lomb Memorial Drive, Rochester, NY 14623, USA}
\affiliation{Space Telescope Science Institute, 3700 San Martin Drive, Baltimore, MD 21218, USA}

\author[0000-0003-2338-5567]{Jes\'{u}s Vega-Ferrero}
\affiliation{Centro de Estudios de F\'{i}sica del Cosmos de Arag\'{o}n (CEFCA), Plaza de San Juan, 1, E-44001 Teruel, Spain}
\affiliation{Departamento de F\'{i}sica Te\'{o}rica, At\'{o}mica y \'{O}ptica, Universidad de Valladolid, 47011 Valladolid, Spain}

\author[0000-0001-6065-7483]{Benjamin J. Weiner}
\affiliation{MMT/Steward Observatory, University of Arizona, 933 N. Cherry St, Tucson, AZ 85721, USA}

\author[0000-0001-9262-9997]{Christopher N. A. Willmer}
\affiliation{Steward Observatory, University of Arizona,
933 North Cherry Avenue, Tucson, AZ 85751, USA}

\author[0000-0002-1333-147X]{Peixin Zhu}

\author[0000-0002-5564-9873]{Eric F.\ Bell}
\affiliation{Department of Astronomy, University of Michigan, 1085 S.\ University Ave., Ann Arbor MI 48109, USA}

\author[0000-0003-3735-1931]{Stijn Wuyts}
\affiliation{Department of Physics, University of Bath, Claverton Down, Bath BA2 7AY, UK}

\author[0000-0002-4884-6756]{Benne W. Holwerda}
\affiliation{Department of Physics, University of Louisville, Natural Science Building 102, 40292 KY Louisville, USA}

\author[0000-0002-9373-3865]{Xin Wang}
\affiliation{School of Astronomy and Space Science, University of Chinese Academy of Sciences (UCAS), Beijing 100049, China}
\affiliation{National Astronomical Observatories, Chinese Academy of Sciences, Beijing 100101, China}
\affiliation{Institute for Frontiers in Astronomy and Astrophysics, Beijing Normal University, Beijing 102206, China}

\author[0000-0002-9593-8274]{Weichen Wang}
\affiliation{Department of Physics, Universita degli Studi di Milano-Bicocca, Piazza della Scienza, 3, Milano I-20126, Italy}

\author[0000-0002-7051-1100]{Jorge A. Zavala}
\affiliation{National Astronomical Observatory of Japan, 2-21-1 Osawa, Mitaka, Tokyo 181-8588, Japan}

%Using this to get the suppressAffiliations to work
\collaboration{1000}{(CEERS collaboration)}

\begin{abstract}
We present the Cosmic Evolution Early Release Science (CEERS) Survey, a 77.2 hour Director's Discretionary Early Release Science Program.  CEERS demonstrates, tests, and validates efficient extragalactic surveys using coordinated, overlapping parallel observations with the {\it JWST} instrument suite, including NIRCam and MIRI imaging, NIRSpec low (R$\sim$100) and medium (R$\sim$1000) resolution spectroscopy, and NIRCam slitless grism (R$\sim$1500) spectroscopy.  CEERS targets the {\it Hubble Space Telescope}-observed region of the Extended Groth Strip (EGS) field, supported by a rich set of multiwavelength data.  
CEERS facilitated immediate community science in both of the extragalactic core \textit{JWST} science drivers ``First Light" and ``Galaxy Assembly," including: 1) The discovery and characterization of  large samples of galaxies at $z \gtrsim$ 10 from $\sim$90 arcmin$^2$ of NIRCam imaging, constraining their abundance and physical nature;  2) Deep spectra of $>$1000 galaxies, including dozens of galaxies at 6 $< z <$ 10, enabling redshift measurements and constraints on the physical conditions of star-formation and black hole growth via line diagnostics; 3) Quantifying the first bulge, bar and disk structures at $z >$ 3; and 4) Characterizing galaxy mid-IR emission with MIRI to study dust-obscured star-formation and supermassive black hole growth at $z \sim$ 1--3.  
As a legacy product for the community, the CEERS team has provided several data releases, accompanied by detailed notes on the data reduction procedures and notebooks to aid in reproducibility.  In addition to an overview of the survey and quality of the data, we provide science highlights from the first two years with CEERS data.
\end{abstract}


\section{Introduction}\label{sec:intro}

The {\it Hubble Space Telescope} ({\it HST}) transformed our understanding of galaxy evolution, from the original Hubble Deep Field pushing the definition of ``high redshift" past $z \sim$ 3 \citep{williams96,madau96,steidel96b,ferguson00,giavalisco02}, to the past decade where the near-infrared capabilities of Hubble's Wide Field Camera 3 pushed to $z \sim$ 10--11 \citep[e.g.][]{ellis13,coe13,oesch16,bouwens19,finkelstein22}. While pushing {\it HST} to its limits enabled observations to reach a time $\sim$500 Myr from the Big Bang, it is clear that this extraordinary 2.4-meter telescope could only reveal the tip of the iceberg of the galaxy population at such an early cosmic epoch. A transformative technological advance was needed to push into the epoch of cosmic dawn at $z \gtrsim$ 10.

This leap is now here with the successful launch and commissioning \citep{rigby23} of {\it JWST}.  {\it JWST}'s 6.5-meter diameter mirror and infrared sensitivity were designed to discover first light in the universe.  The large mirror size is necessary to capture light emanating from faint, distant galaxies, while the infrared sensitivity is necessary to capture highly redshifted light from these rapidly receding galaxies.  {\it JWST} thus has the capability to not only probe the full galaxy population at $z \sim$ 10, but to push to higher redshifts and answer several outstanding questions in early-universe astrophysics.  When did the dark ages end?  How abundant were early galaxies?  Is the physics dominating star formation the same at early times, or do changing physical conditions lead to changes in, for example, the star-formation efficiency or the initial mass function?

In addition to {\it JWST}'s ability to discover early galaxies in this epoch with NIRCam \citep{rieke23} imaging, its spectroscopic capabilities are also revolutionary, with orders-of-magnitude gain in spectroscopic sensitivity at 2.5--5$\mu$m, precisely where early galaxies have strong rest-frame optical line emission.  Multi-object spectroscopic surveys with both the NIRSpec \citep{jakobsen22} micro-shutter array (MSA) and NIRCam wide-field slitless spectrograph allow measurement of, for the first time, not only large numbers of spectroscopic redshifts in the epoch of reionization, but the first direct constraints on ionizing conditions and chemical evolution in these early galaxies.  Finally, the MIRI \citep{wright23} imager probes 5--25 $\mu$m, sensitive to dust emission at $z \sim$ 1--3 and rest-frame optical stellar continuum emission in the epoch of reionization, dramatically improving constraints on the stellar masses and star-formation rates (SFRs) of early galaxies \citep[e.g.][]{papovich23}, which may have been significantly biased by previous measurements probing solely the rest-frame ultraviolet (UV).

The potential for {\it JWST} to revolutionize all areas of astrophysics led to the creation of the Director's Discretionary (DD) Early Release Science (ERS) program.  This program dedicated $\sim$500 hours of the earliest science time in the first year of the {\it JWST} mission to obtaining data spanning all areas of astronomy, from the Solar System to the early universe.  Here we describe one of these 13 approved programs -- the Cosmic Evolution Early Release Science (CEERS) Survey.  The CEERS survey combines four modes of {\it JWST} operations (NIRcam imaging, NIRCam grism spectroscopy, NIRSpec MSA spectroscopy, and MIRI imaging) to validate efficient parallel survey operations with {\it JWST} and address key open questions in galaxy formation, while also allowing a variety of investigations into galaxies over the epoch 0.5 $< z <$ 12 (and beyond), setting the stage for the full {\it JWST} mission.  

CEERS builds on past community surveys from space-based observatories. 
 Since their inception with the Hubble Deep Field, Hubble Ultra-Deep Field \citep{beckwith06}, and the Great Observatory Origins Deep Survey \citep[GOODS; ][]{dickinson03,giavalisco04}, public surveys have demonstrated their power as effective science multipliers by attracting a vast array of highly complementary, high-quality data sets from a broad array of facilities in a number of carefully selected ``Legacy Fields". By being 
immediately available to the community for scientific exploitation and further investigation, these panchromatic data sets are effectively acting as powerful science aggregators that are primary targets when new facilities such as {\it JWST} become available. In distant galaxy studies, five of these Legacy Fields stand out: GOODS North and South (with the Hubble Deep Field and Hubble Ultra Deep Field embedded inside, respectively), UDS, COSMOS, and EGS.  These five fields were deeply observed by {\it HST} with the CANDELS program \citep{grogin11,koekemoer11}, and unsurprisingly, all five of these fields are currently being targeted by {\it JWST}, with CEERS targetting the Extended Groth Strip (EGS) field.

While large {\it HST} survey programs have been among the most successful in terms of publications per orbit, {\it JWST} is a more complicated observatory, with different restrictions on observability, parallel mode usage, and exposure timing constraints.  CEERS was designed to validate several modes of {\it JWST} survey observations, including multiple
parallel modes, rapidly providing valuable data to the community, and setting the stage for efficient surveys
to be done in early {\it JWST} cycles.  

\begin{figure*}[!t]
\epsscale{1.15}
\plotone{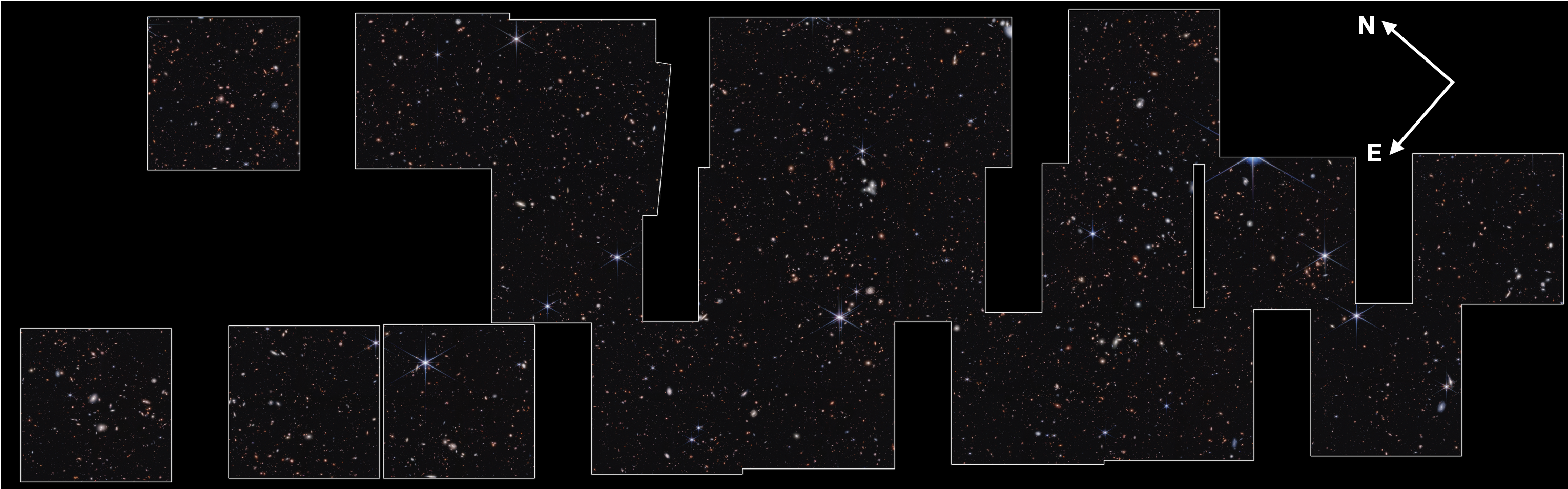}
\caption{A color image of the CEERS NIRCam imaging, made using all seven filters.  These data cover $\sim$90 arcmin$^2$ in the CANDELS EGS Field.  These data were obtained in parallel to prime NIRSpec and MIRI observations.   The module gaps (and short-wavelength chip gaps) were not filled due to the limited time available for ERS programs, leading to the unique imaging footprint shown.  Figure credit: Alyssa Pagan (STScI).  This image is available for download with this news release: https://webbtelescope.org/contents/news-releases/2023/news-2023-114}
\label{fig:fieldimage}
\end{figure*}

In this survey overview paper, we present the motivation and technical details behind the CEERS survey.  In \S 2 we highlight the science justification beind CEERS, while in \S 3 we describe the CEERS observing strategy, including justification for the various choices in survey design.  Our datasets are described in \S 4, while in \S 5 we summarize science highlights from papers.  In \S 6 we discuss lessons learned from this program in the context of future public survey endeavors.  We illustrate the NIRCam portion of CEERS with a full-color mosaic of our completed NIRCam imaging in Figure~\ref{fig:fieldimage}. Throughout this paper all magnitudes are reported on the AB system \citep{oke83}.

\section{Scientific Motivation}

Between $z \gtrsim$ 10 and today, galaxies underwent dramatic transformations. Gas fell into dark matter halos, where it cooled, condensed and started to form stars, building stellar populations within galaxies, reaching peak activity at $z \sim$ 1--3 \citep{madau14}, and enhancing their metal and dust content. Their central supermassive black holes (SMBHs) grew, leading to a relationship between SMBH and stellar mass \citep{kormendy13,greene20}. This growth was accompanied by changes in the physical structures of galaxies \citep[e.g.][]{genzel17}, as they grew their disks and became increasingly bulge-dominated. Galaxies also grew through mergers, further enhancing star formation and black hole growth, while driving morphological changes. In this section, we describe the scientific motivation behind the CEERS program, designed to improve our understanding of the key evolutionary pathways that have built-up today's galaxies.

\subsection{First Light in the Universe}

\subsubsection{What are the limits on the epoch of first galaxy formation?}

The shape of the rest-frame UV luminosity function constrains the
relative importance of the physical processes governing the conversion
of gas into stars. These processes depend on numerous factors,
including density, metallicity, magnetic field strength, turbulence,
and include a variety of feedback mechanisms.  At the highest redshifts we
have almost no empirical constraints on these processes.  While
locally the star-formation efficiency is a few percent per free-fall
time \citep[e.g.][]{kennicutt98}, this is likely very different in the early Universe. 
Lower metallicities could reduce cooling efficiency and delay molecular hydrogen formation and thus subsequent star-formation \citep[e.g.][]{krumholz12}, while higher gas densities, similar to those seen only in extreme starbursts in the local
universe, could lead to a steeper dependence of the SFR on the gas density \citep[e.g.][]{krumholz09,ostriker11}, and a thus
\emph{more} efficient conversion of gas into stars \citep{somerville15b}. 

It is unclear what the net effect of these different factors will be, and whether the process of
converting dense gas into stars at early times was more or less
efficient than in the local Universe, leaving one of the most
fundamental ingredients in galaxy formation models highly uncertain.  Unsurprisingly, model predictions spanned a wide range, with models which were able to reproduce the observed UV luminosity functions at $z \lesssim$ 6 diverging increasingly at higher redshift, showing a more than 1 dex difference in the prediction for the abundance of $M_{UV}<-$19 galaxies at $z>$11 \citep[e.g.][]{tacchella13,mason15,behroozi15,tacchella18,behroozi20,gnedin16,xu16,wilkins17,ma18,yung19a, yung20b}.  

When the call for {\it JWST} ERS proposals was released in 2016, the field of high-redshift galaxy evolution had its eye on $z \sim$ 6 -- 8.  Years of investment in surveys with {\it HST}s WFC3/IR, such as CANDELS \citep{grogin11,koekemoer11} and the recently completed Hubble Frontier Fields \citep{lotz17} had uncovered samples of 1000's of galaxies in this epoch.  However, {\it HST}/WFC3 only scratched the surface of the $z=$9-10 universe due to a combination of extremely small samples close to the
image detection limits, with many candidate galaxies detected in only a single
filter \citep[e.g.][]{ellis13,coe13,oesch13,oesch14,bouwens15,mcleod15,morishita18,bouwens21,finkelstein22}.   {\it JWST} changed the game as NIRCam can select fainter, higher--redshift galaxies (to $z \sim$ 20) than possible with {\it HST}, allowing robust construction of the UV luminosity function at $z \sim$ 10--12, with possible constraints to $z \sim$ 15 and beyond.

\subsubsection{Chemical Enrichment in the Early Universe}

As we discover galaxies closer and closer to the Big Bang, at some point we should begin to witness the periods during which galaxies are very young, likely having formed no more than a few generations of stars.  These galaxies will be characterized by extremely low metallicities.  Modern simulations now show that Pop III star formation in early mini-halos at $z \sim$ 15-20 was likely very efficient at polluting the IGM with metals, and so we might \emph{never} expect to see metal-free star formation \citep[e.g.][]{jaacks18b}.  The consequences for subsequent star formation are critical — if all dense gas in halos is rapidly enriched beyond the critical metallicity (needed to transition to a Population II initial mass function; Z $\sim$ 10$^{-4}$ Z\sol), both the stellar initial mass function and stellar photospheric temperatures will likely not be dramatically different than that seen in low-metallicity environments in the local universe.  If the opposite is true, and fairly massive metal-free stars can form down to even $z \sim$ 10, it will provide a distinct boost in the typical hardness of stellar spectra, increasing the production efficiency of ionizing photons, with consequences on the ability of stellar light to reionize the IGM.

Constraints on the typical metallicities of early galaxies are thus of intense interest.  A straightforward measure of the physical properties of the stars is available via the UV spectral slope $\beta$ (f$_{\lambda} \propto \lambda^{\beta}$; \citealt{calzetti94}), as it can be measured with just a few photometric points.  Constraining this quantity was thus a major focus when the first {\it HST}/WFC3 observations allowed the study of the spectral slope of $z =$ 6–8 galaxies, with studies finding that presently observable galaxies are consistent with somewhat low, but non-zero, metallicities without much dust obscuration \citep[e.g.][]{finkelstein12a,dunlop13,bouwens14}.  It was ancitipated that pushing to higher redshifts with \emph{JWST} would transform this field in a variety of ways.  First, previous measures of galaxies at $z =$ 6–8 were restricted to just one or two colors, which introduced strong statistical and systematic errors \citep{finkelstein12a}.  CEERS NIRCam imaging allow the study of the rest-UV colors of $z \sim$ 6–15 galaxies using 2--4 colors.  Coupled with the sensitivity of CEERS this allows much more robust measures of $\beta$, as well as probing a larger dynamic range in galaxy luminosity.

\subsubsection{A Large Spectroscopic Sample in the Epoch of Reionization}

Photometric redshifts are susceptible to both statistical uncertainties in the redshift ($\sigma_z \sim$ 0.2--0.5), which propagate through to key quantities such as the luminosity, stellar mass and SFR, and potential contamination by lower-redshift galaxies or cool stars \citep[][and references therein]{finkelstein16}.  These uncertainties must be calibrated with spectroscopic redshifts to allow a meaningful comparison to models.  Previously, the spectroscopic tracer of choice has been Ly$\alpha$.  While readily detectable from the majority ($\sim$60\%) of observed galaxies at $z \sim$ 5–6 \citep[e.g.,][]{stark11}, the shift to the near-infrared coupled with reduced detectability due to an increasingly neutral IGM makes Ly$\alpha$ emission harder to detect from galaxies \citep[e.g.][]{pentericci14,larson22}.

{\it JWST} near-infrared spectroscopy transforms this field by revealing rest-optical emission lines in $z >$ 6 galaxies for the first time, tracing redshifts via expectedly strong \citep[e.g.][]{shim11,finkelstein13,stark13,smit15}  [O\,{\sc iii}] (to $z <$ 9.5) and/or H$\alpha$ (to $z <$ 6.7).  This will make spectroscopic redshifts ``trivial" for most galaxies, removing uncertainties inherent in photometric redshifts.  For more distant galaxies, spectroscopic confirmation is still achievable via prism observations of the Ly$\alpha$ continuum break, or weaker rest-UV lines such as C\,{\sc iii}] or C\,{\sc iv} (or even Ly$\alpha$).

\begin{figure*}
\includegraphics[width=\textwidth]{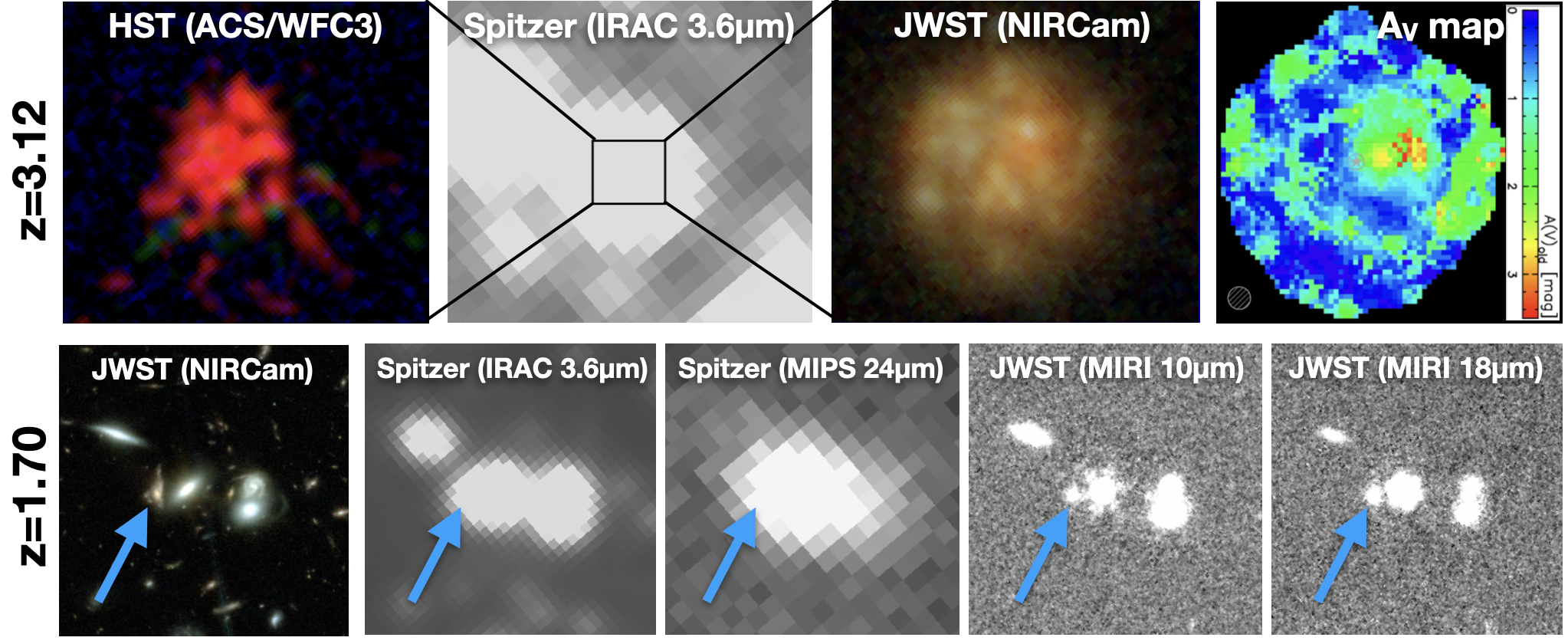}
\caption{Images of two galaxies in the CEERS field, demonstrating the resolution gain over previous observatories.  The top row shows a $z_{spec} =$ 3.13 galaxy, which due to its red color was only well-detected by {\it HST} in the F160W filter, though was clearly detected by {\it Spitzer}/IRAC.  The improvement with NIRCam is startling, with the higher resolution ($>$10$\times$ better at 3.6$\mu$m compared to IRAC) revealing a clear spiral galaxy.  This exquisite resolution allows investigation into this galaxy's detailed structure, as illustrated by the dust attenuation map in the right-most panel \citep{perezgonzalez23}.  The bottom row shows a $z =$ 1.70 galaxy.  The NIRCam image reveals it to have a red color, potentially indicative of significant obscured star-formation.  However, the poor resolution of {\it Spitzer} at 3.6 (PSF FWHM $\sim$1.7\arcs) and 24 $\mu$m ($\sim$6\arcs) prohibited investigation into the dust emission due to crowding by nearby objects.  Our CEERS MIRI images clearly reveal emission from this source at 10 (0.32\arcs) and 18 $\mu$m (0.59\arcs).}
\label{fig:resolutionfig}
\end{figure*}

\subsection{Galaxy Assembly: The growth of galaxies and their SMBHs from \boldmath{$z \sim$7} to 1}

\subsubsection{Evolution of ISM Properties}
Prior to the launch of \textit{JWST}, spectroscopy of $z >$1 galaxies revealed very different emission-line properties compared to local $\HII$ regions, implying strong evolution in ionization, gas density, metallicity, stellar winds, binary fractions, and/or AGN emission \citep[e.g.][]{kewley16,stark14,shapley16,steidel18}.  \textit{JWST}'s ability to measure rest-UV and rest-optical emission lines at $z >$ 3 is transformative, constraining evolution in gas-phase metallicity, ionization, and pressure \citep[][and references therein]{kewley19}.  NIRSpec and NIRCam spectra cover a broad suite of emission lines at 1--5$\mu$m.  This includes rest-optical lines sensitive to ionization and metallicity (e.g., BPT, R23, $\OIII\lambda4363$, $\NeV\lambda3426$), and rest-UV lines (e.g., $\CIV\lambda1550$, $\CIII\lambda1909$, $\HeII\lambda1640$) as indicators of AGN/shocks, (binary) stellar populations, and ionizing photon production \citep{gutkin16, jaskot16, steidel16, stark17}.  NIRCam slitless grism spectra can map spatially resolved star formation, shocks, and nuclear AGN \citep{trump11, nelson16}.  Combining spectroscopy with the deep NIRCam and MIRI imaging enables studies of correlations between galaxy ISM properties and galaxy mass, SFR, and morphology.

\subsubsection{Stellar Mass Growth Across Cosmic Dawn and Cosmic Noon}
Observations of massive galaxies at high redshifts provide a crucial testing ground for theories of early, rapid galaxy growth and stellar or AGN feedback.  Prior to {\it JWST}, the census of massive galaxies at $z >$3 was highly uncertain:  these galaxies are typically faint in the rest-UV, and red due to dust and age.  Very few photo--$z$ selected massive galaxies had spectroscopic confirmation \citep[e.g.][]{marsan17,glazebrook17}.   {\it JWST} near/mid-IR photometry enables robust stellar mass measurements at the highest redshifts. The combination of CEERS NIRCam$+$MIRI imaging has significantly reduced degeneracies due to age, dust and nebular emission in stellar population modeling, allowing robust stellar mass estimates for high-redshift galaxies. CEERS spectroscopy has allowed direct measures of the emission-line contribution to broadband fluxes, enabling statistical corrections for the full
high-redshift population. CEERS NIRSpec observations also target $\sim$40 massive-galaxy candidates at $z\!>$3
demonstrating the ability to measure redshifts and stellar populations via absorption lines.

\subsubsection{The Structural Evolution of Galaxies}

While \textit{HST} has greatly enhanced our understanding of galaxy structure and morphology, particularly at moderate redshifts ($z\sim1$), the details of how these structures are first put into place and the precise evolutionary pathways these early galaxies then follow over the age of the universe still elude us \citep{wuyts11,lang14,tacchella15}. For example, how the first massive disks in the early universe formed, how and when the formation of the first bulges took place, and the physical mechanisms responsible for driving the fueling and quenching of star formation in galaxies (a process that appears to be inextricably linked to morphological transformation for some galaxies) are still largely unsolved questions. The combination of NIRCam and NIRSpec observations using \textit{JWST} are enabling us to address these questions for the earliest galaxies and quantify how these mechanisms have evolved over cosmic time.  

The CEERS NIRCam resolution corresponds to $\lesssim$~1~kpc at 3.6~$\mu$m (a factor of 8 improvement over \textit{Spitzer}/IRAC, see Figure~\ref{fig:resolutionfig}), and enables spatially resolved measurements of galaxies in the rest-optical/IR at 1$<$$z$$<$8, well-beyond the volume accessible to \textit{HST}. The photometry will be used to build SEDs, measure photometric redshifts, and estimate physical parameters like stellar masses and star formation rates. The imaging will be used to classify objects morphologically using visual, parametric (e.g., Sersic parameters via Galfit), and non-parametric (e.g., concentration, asymmetry, Gini, M20, etc.) measures, from which objects with regular (bulges, disks) and irregular or perturbed morphologies will be identified and used to examine how their relative proportions evolve from very high redshifts to ``cosmic noon." This will enable us to quantify when the first massive disks and bulges formed and place constraints on the role of interactions and mergers in the morphological transformation of galaxies.

For example, a key characteristic of quiescent galaxies at $z\sim2$ is their compact size \citep[e.g.][]{daddi05,trujillo07,vandokkum08,cassata11}. Passive galaxies with stellar masses of $M\sim10^{11}$ $M_{\odot}$ are $\sim4$ times smaller and two orders of magnitude more dense at $z=2$ than they are at $z=0$ \citep{vanderwel14,bezanson09}. These quiescent galaxies are the likely progenitors of the most massive, early-type galaxies found locally \citep[e.g.][]{hopkins09,vandokkum14}, and studying the processes that give rise to this population is central to understanding the origins of their local counterparts. There is currently much debate as to how massive, compact galaxies formed in the early universe. A prevailing scenario posits that they are the descendants of larger, more extended star-forming galaxies that underwent a compaction phase at $z=2-3$ as a result of gas-rich, dissipative processes, such as gas-rich mergers or violent disk instabilities 
\citep{barro13,barro14,zolotov15,tacchella16}.  Alternatively, the dense cores of these galaxies may have formed in situ at even higher redshifts ($z-3-5$), when all galaxies were denser, and the resulting galaxies remained compact until $z\sim2$ \citep{wellons15,lilly16,williams17}. With CEERS NIRCam imaging, we directly test the early formation model by determining if a large population of massive, compact SFGs is already in place at $z=3-5$. 

The spatial resolution, sensitivity, and long-wavelength coverage of NIRCam also enables resolved SED fitting in order to measure the spatial distribution of stellar mass, star formation rate, and dust throughout individual galaxies \citep[e.g.][]{wuyts12,hemmati14,hemmati15,tacchella15,jung17} at higher redshift than has previously been possible. This will provide an independent selection of quiescent galaxies (low sSFR in the center) and, combined with the morphological measurements described above, will constrain the importance of different quenching mechanisms (outside-in vs. inside-out).

\subsubsection{Dust-Obscured Star-Formation and SMBH Accretion}

\textit{JWST} observations are now transforming our census of the dust--attenuated phases of star formation in galaxies and AGN at $z \sim$1--3, when such objects were most active \citep{madau14}. 
\textit{JWST} enables the characterization of the mid-IR emission of these galaxies, where a significant amount of the energy from star-forming galaxies is emitted by polycyclic aromatic hydrocarbons \citep[PAHs, e.g., the 6.2, 7.7, and 8.3\; $\mu$m complexes;][]{smith07}, or from toroidal hot dust in AGN \citep[e.g.,][]{kirkpatrick17}. 
MIRI observations can easily detect IR emission from galaxies and obscured AGN a factor 10 fainter than possible with previous IR surveys, from $\sim$$10^8$\,$L_\odot$ at $z\sim 1$ to $\sim$$10^{9}$~$L_\odot$ at $z\sim 2$, during the period when the SFR density peaked \citep{madau14}.

\begin{figure*}[!t]
\epsscale{1.0}
\plotone{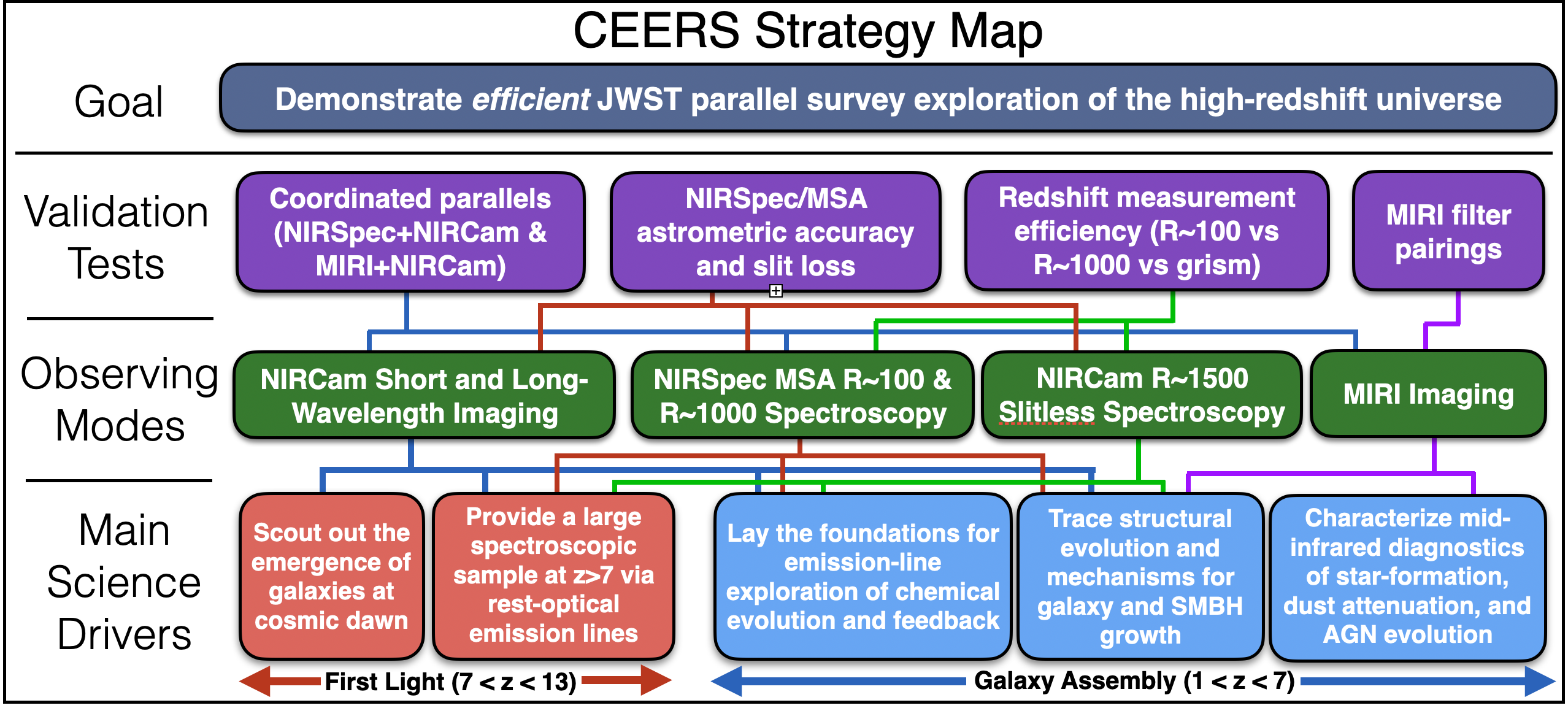}
\caption{This diagram outlines the strategy behind the CEERS survey.  The strategy flows down from the overall goal, through validation tests, to determine the observing modes, each of which are linked to one or more of our main science drivers.}
\label{fig:barbados}
\end{figure*}

 At $z \sim$1--3 MIRI coverage from 7.7--21\; $\mu$m enables a wide range of science. First and foremost, MIRI has a 10$\times$ improvement in angular resolution over {\it Spitzer}/MIPS, allowing for unambiguous source identification. The close spacing of MIRI filters allows for the characterization of the mid-IR emission, as either star-forming or AGN, through identification of PAH features \citep{kirkpatrick17}. MIRI will also allow for the calculation of SFRs from single band estimators tied to PAH luminosities \citep[e.g.,][]{shipley16}. Robust coverage of the mid-infrared emission, combined with \textit{HST} coverage, will also allow for the study of dust emission properties, PAH feature strength, and the attenuation of UV-optical emission. It will enable the dust attenuation law at $z\sim1-3$ to be be constrained with higher precision than previously possible \citep[e.g.][]{buat11,scoville14,salmon16}.

\subsubsection{Black Hole Growth Across Cosmic Time}

\textit{JWST} enables several studies into the relationship between SMBH growth and galaxy evolution out to $z>7$. Spectroscopy and imaging can be used to identify AGN in different evolutionary phases using line ionization diagnostics, mid-IR colors, and broad-line detections.
The NIRCam imaging will provide sub-kpc spatial resolution at 1-4$\mu$m and enable the study of AGN host morphologies in the rest-frame optical at $z>3$ for the first time.  This will enable constraints on proto-bulge growth, and it will determine the mechanisms fueling SMBH growth during the era when the AGN-galaxy connection is being established. The broad wavelength coverage of the NIRSpec observations (1-5$\mu$m) will provide a wide suite of independent line ratios that can be used to robustly diagnose whether photoionization in galaxies is dominated by young stars or an AGN.  The NIRSpec data will be sensitive to high ionization lines (N\,{\sc v}, He\,{\sc ii}, C\,{\sc iv}, Ne\,{\sc v}) from growing SMBHs at $z>7$, as tentatively seen in a few high-$z$ Ly$\alpha$-emitters \citep{Tilvi16}. The discovery of a large population of such sources would suggest that weak AGN may help clear the neutral hydrogen that surrounds galaxies during the epoch of reionization.  The resolution ($R$) of the available spectroscopic gratings also enable the measurements of emission-line velocity widths, permitting the first measurements of AGN driven outflows and feedback at $z>3$ and enabling the detection of lower-luminosity broad-line AGN out to $z>7$.  These sources would be among the highest redshift AGN ever discovered.

\section{CEERS Observing Strategy}

Making progress on these science questions early in Cycle 1 required a modest-sized survey involving multiple {\it JWST} instruments.  CEERS aimed to accomplish this by providing the needed data using these observing modes:
\begin{itemize}
\item NIRCam short and
  long-wavelength imaging from 1--5$\mu$m
\item MIRI imaging from 5--21$\mu$m 
\item NIRSpec multi-slit R$\sim$100 and 1000 spectroscopy from 1--5$\mu$m 
\item NIRCam slitless grism spectroscopy (3--4$\mu$m)
\end{itemize}
CEERS is comprised of ten pointings of NIRCam imaging, eight pointings of MIRI imaging, four pointings of NIRCam wide field slitless spectroscopy, and eight pointings of NIRSpec multi-object spectroscopy. The pointings were obtained as pairs of coordinated parallel observations (NIRCam imaging in parallel to both NIRSpec and MIRI; MIRI in parallel to NIRCam grism) designed to maximize the overlap between the different instrument footprints. 

These observations provide an early demonstration of an efficient {\it JWST} parallel survey, both showcasing the parallel instrument pairs as viable survey modes, and providing {\it JWST} data early in the mission to begin to address the key science goals. The wealth of imaging and spectroscopic data are allowing a variety of blank-field extragalactic studies, making the CEERS data of immense interest to a large cross-section of the astronomical community and serving as a pathfinder for scientific programs in future cycles.  The flowdown from our primary science goal, through the validation tests, to observing modes and science drivers is shown in Figure~\ref{fig:barbados}\footnote{Modeled after the strategic map for the nation of Barbados: \url{https://www.strategymapexample.com/example-of-the-strategy-map-that-can-be-implemented-for-international-company.htm}}. Section 3.1 describes the field choice, while Sections 3.2--3.5 describe the observing plans for specific instruments.  Section 3.6 describes the observing layout and scheduling plan, with the survey layout as implemented presented in Figure~\ref{fig:layout}.  We summarize the instrument coverage and pointings centers in Table~\ref{tab:obs}.

\begin{deluxetable*}{lccccc}
\tablecaption{CEERS Observation Summary\label{tab:obs}}
\tabletypesize{\small}
\tablehead{
\colhead{Pointing} & \colhead{NIRCam Center} & \colhead{MIRI Center} &
\colhead{NIRSpec Center} & \colhead{Obs Date} & \colhead{V3 PA}
}
\startdata
1                  & 14:19:56.2 +52:58:38.8 & 14:20:38.9 +53:03:04.6 & \nodata                & 21 Jun 2022    & 130.8 \\
2                  & 14:19:34.8 +52:54:50.3 & 14:20:17.4 +52:59:16.2 & \nodata                & 21-22,28 Jun 2022 & 130.8 \\
3                  & 14:19:12.7 +52:51:03.5 & 14:19:55.2 +52:55:29.4 & \nodata                & 22 Jun 2022    & 130.8 \\
4                  & 14:19:02.1 +52:46:05.4 & \nodata                & 14:19:24.1 +52:51:35.8 & 21 Dec 2022    & 310.8 \\
5\tablenotemark{a} & 14:19:46.0 +52:53:37.6 & 14:19:05.2 +52:49:27.5 & 14:20:08.1 +52:59:06.6 & 21,24 Dec 2022 & 310.8 \\
6                  & 14:19:25.2 +52:49:56.0 & 14:20:07.8 +52:54:21.8 & \nodata                & 22 Jun 2022    & 130.8 \\
7\tablenotemark{a} & 14:20:25.7 +52:57:12.3 & 14:19:45.1 +52:53:03.8 & 14:20:47.8 +53:02:41.2 & 21,24 Dec 2022 & 310.8 \\
8\tablenotemark{a} & 14:20:03.5 +52:53:21.9 & 14:19:22.6 +52:49:14.1 & 14:20:25.5 +52:58:50.9 & 20-21 Dec 2022 & 310.8 \\
9\tablenotemark{a} & 14:19:41.1 +52:49:35.9 & 14:19:00.2 +52:45:28.0 & 14:20:03.1 +52:55:04.9 & 20-22,24-25 Dec 2022 & 310.8 \\
10                 & 14:19:18.9 +52:45:48.3 & \nodata                & 14:19:40.9 +52:51:19.2 & 22,25 Dec 2022 & 310.8 \\
11                 & \nodata                & \nodata                & 14:19:37.0 +52:50:34.1 & 9-10 Feb 2023  & 262.9 \\
12                 & \nodata                & \nodata                & 14:19:06.8 +52:44:49.5 & 10 Feb 2023    & 263.2 \\
DDT 2750           & 14:20:15.5 +52:52:22.0 & \nodata                & 14:19:36.9 +52:54:58.9 & 24-25 Mar 2023 & 213.7 \\
\enddata
\tablenotetext{a}{CEERS pointings 5, 7, 8 and 9 were observed with both prime NIRSpec MOS + parallel NIRCam imaging and prime NIRCam WFSS + parallel MIRI imaging. The prime and parallel NIRCam observations for each pointing target the same footprint.
}
\end{deluxetable*}

\subsection{Field Choice}

Considerations when choosing the position in the sky for CEERS included low Galactic extinction, low zodiacal background, wide-field \emph{HST} pre-imaging for NIRSpec MSA planning, and extensive multi-wavelength imaging and spectroscopy to maximize scientific utility.  The five CANDELS \citep{grogin11,koekemoer11} fields broadly meet these requirements.  Of these fields, the GOODS North and GOODS South fields were already scheduled to be targeted by the JADES Guaranteed Time Observation program \citep{eisenstein24}, and were thus unavailable for a wide-field survey such as CEERS due to the {\it JWST} duplication policy.

Of the three remaining fields, we selected the EGS as ideal for a distant-universe \emph{JWST} ERS program.  The geometry and location of the EGS permit efficient coordinated \emph{JWST} parallels and long observability windows early in Cycle 1.  Specifically, the observable V3PA of the EGS field combined with the \textit{JWST} focal plane layout provide an optimal self-overlap of CEERS \textit{JWST} observations, allowing scientific investigations using, for example, \textit{JWST} NIRCam and MIRI imaging and NIRSpec spectroscopy for the same source.  Conversely, the combination of the position angle of the COSMOS and UDS fields on the sky with their observable V3PAs provides a sub-optimal overlap of \textit{JWST} instruments with \textit{HST} imaging (as well as for \textit{JWST} self-overlap).  

The EGS field has high ecliptic latitude compared to the COSMOS and UDS fields, resulting in lower and more constant zodiacal background levels, yielding NIRCam imaging depths up to 0.2 mag deeper compared to minimum background levels in those other fields.  The EGS field also has the largest number of bright ($m <$ 26.5) {\it HST}-identified $z \gtrsim$ 9 candidate galaxies \citep[e.g.][]{bouwens19,finkelstein22}, including a significant overdensity at $z =$ 8.7 \citep{larson22,zitrin15,whitler23}.  We note that both COSMOS and UDS were also observed in Cycle 1 GO programs (Primer [PI Dunlop] and COSMOS-Web [PIs Kartaltepe \& Casey, \citealt{casey23a}]), meaning that all five CANDELS fields now have significant {\it JWST} investment.

%% NIRCam exposure table 
\begin{deluxetable*}{lcccccccc}
\tablecaption{NIRCam Exposure Times\label{tab:nircam_exptimes}}
\tabletypesize{\small}
\tablehead{
\colhead{Instrument Pointing} & \colhead{F115W} & \colhead{F150W} & \colhead{F200W}
& \colhead{F277W} & \colhead{F356W} & \colhead{F410M} & \colhead{F444W} & \colhead{WFSS F356W}
}
\startdata
NIRCam1  & 6184.4 & 3092.2 & 2834.5 & 3092.2 & 3092.2 & 3092.2 & 2834.5 & \nodata  \\
NIRCam2  & 6184.4 & 3092.2 & 2834.5 & 3092.2 & 3092.2 & 3092.2 & 2834.5 & \nodata  \\
NIRCam3  & 5669.0 & 2834.5 & 2834.5 & 2834.5 & 2834.5 & 2834.5 & 2834.5 & \nodata  \\
NIRCam4  & 5669.0 & 2834.5 & 2834.5 & 2834.5 & 2834.5 & 2834.5 & 2834.5 & \nodata  \\
NIRCam5  & 10028.2 & 2834.5 & 2834.5 & 2834.5 & 4702.7 & 2834.5 & 2834.5 & 2490.9 \\
NIRCam6  & 5669.0 & 2834.5 & 2834.5 & 2834.5 & 2834.5 & 2834.5 & 2834.5 & \nodata  \\
NIRCam7  & 10028.2 & 2834.5 & 2834.5 & 2834.5 & 4702.7 & 2834.5 & 2834.5 & 2490.9 \\
NIRCam8  & 10028.2 & 2834.5 & 2834.5 & 2834.5 & 4702.7 & 2834.5 & 2834.5 & 2490.9 \\
NIRCam9  & 12862.7 & 2834.5 & 2834.5 & 2834.5 & 4702.7 & 2834.5 & 5669.0 & 2490.9 \\
NIRCam10  & 5669.0 & 2834.5 & 2834.5 & 2834.5 & 2834.5 & 2834.5 & 2834.5 & \nodata  \\
DDT2750  & 6345.4 & 5701.2 & 5701.2 & 6345.4 & 5701.2 & \nodata & 5701.2 & \nodata  \\
\enddata
\tablecomments{Exposure time (s) in each NIRCam pointing and filter. Exposure times in F115W and F356W for pointings 5, 7, 8 and 9 include the direct and out-of-field imaging obtained as part of the WFSS observations. The WFSS observations are performed with 1245.5 s each of GRISMR and GRISMC.}
\end{deluxetable*}

\subsection{NIRCam Survey}
The CEERS observing strategy is set by our primary science goal:  detecting a large sample of $z\sim$9--13 galaxies to explore the evolution of the UV luminosity function to $z \sim$ 10 and beyond.  When designing the survey, we conducted a depth--versus--area trade study using pre-launch model predictions, and found that detecting galaxies to 
$M_{UV}\!=-$19.5 over 90 arcmin$^{2}$ maximizes the number of $z\!>$9 sources (as opposed to, for example, going deeper in a single pointing in the same total amount of observing time). This led to our NIRCam observing plan of 10 pointings.  Covering larger areas in fixed time led to diminishing returns in terms of number of $z\!>$9 sources; while larger areas would reduce the number of MIRI and/or NIRSpec observations. 

We observed with seven filters in all 10 NIRCam pointings, pairing the SWC+LWC filters: F115W$+$F277W, F115W$+$F356W and F150W$+$F410M, and F200W$+$F444W.  We observe them in this order to ensure that persistence from previous observations does not mimic a Lyman break (as it would if we observed the redder filters first).  We reach our desired sensitivity in 2835 sec of total integration time (3 dithers of single-integration exposures of 9 groups each in MEDIUM8 readout mode; we choose MEDIUM8 over DEEP8 to achieve more groups [9 vs 5] for improved ramp fitting and cosmic ray identification).  Pre-launch we estimated that this would achieve $>$10$\sigma$ depths for unresolved galaxies and 5$\sigma$ depths for {\it resolved} galaxies (assuming a half-light radius of r$_h$$\sim$$0\farcs1$; \citealt{shibuya16}) at 9$<$$z$$<$13 with rest-frame $M_{UV} \leq$--19.5 ($m$=28.0-28.2). With a higher background, F444W is $\sim$0.3 mag shallower, sufficient for stellar mass determination.   As F115W is our dropout band for our primary science goal of finding $z > 9$ galaxies, we require additional imaging to allow Ly$\alpha$-break selection to the limit of the F150W image, thus we add a second 2835 sec set of integrations in this filter.  In 4 fields, these additional F115W exposures are paired with the long-wavelength grism observations; in the other 6 fields, these are paired with additional long-wavelength imaging exposures.

There are a few exceptions to our nominal 9-group imaging exposures.  The first is the blue-channel imaging alongside our grism exposures, which we elect to do in F115W to further increase the depth of the dropout band in these four fields.  Therefore, in our grism fields, NIRCam fields 5, 7, 8 and 9 (see Figure~\ref{fig:layout}), we must follow a permitted grism dither pattern.   We use a four-point NIRCam+MIRI dither pattern, which drives us to a SHALLOW4 read with 6 groups per integration (total integration time of 2490 sec, close to our desired 2835 sec).  We note that an additional 622 sec of F115W+F356W imaging are obtained with the two direct images (one for the row and one for the column grism).  Another 1244 sec are obtained with the out-of-field images (622 sec x 2 grism), though these, by definition, do not completely overlap our fiducial pointing.  The second exception is for the NIRCam pointings in parallel to MIRI pointings 1 and 2.  In these MIRI pointings, we desire to fully sample the IR SED, which drives us toward $>$3 filters.  As MIRI filter changes are not permitted during NIRCam exposures, we reduced the number of groups in each NIRCam exposure from 9 groups to 5 groups and instead used twice the number of exposures. In these pointings we achieve a limiting magnitude $<$0.1 mag different from the rest of the mosaic.
We summarize the NIRCam exposure times for each pointing and filter in Table~\ref{tab:nircam_exptimes}.

%% MIRI exposure table
\begin{deluxetable*}{lccccccc}
\tablecaption{MIRI Exposure Times\label{tab:miri_exptimes}}
\tabletypesize{\small}
\tablehead{
\colhead{Instrument Pointing} & \colhead{F560W} & \colhead{F770W} & \colhead{F1000W}
& \colhead{F1280W} & \colhead{F1500W} & \colhead{F1800W} & \colhead{F2100W}
}
\startdata
MIRI1  & \nodata & 1648.4  & 1673.3  & 1673.3  & 1673.3  & 1698.3   & 4811.9  \\
MIRI2  & \nodata & 1648.4  & 1673.3  & 1673.3  & 1673.3  & 1698.3   & 7883.9  \\
MIRI3  & 2938.5  & 8815.4  & \nodata & \nodata & \nodata & \nodata  & \nodata \\
MIRI5  & \nodata & \nodata & 1243.2  & 932.4   & 932.4   & 1243.3    & \nodata \\
MIRI6  & 2938.5  & 8815.4  & \nodata & \nodata & \nodata & \nodata  & \nodata \\
MIRI7  & 1433.4  & 2580.1  & \nodata & \nodata & \nodata & \nodata  & \nodata \\
MIRI8  & \nodata & \nodata & 1243.2  & 932.4   & 932.4   & 1243.3   & \nodata \\
MIRI9  & 1433.4  & 2580.1  & \nodata & \nodata & \nodata & \nodata  & \nodata \\
\enddata
\tablecomments{Exposure time (s) in each MIRI pointing and filter. MIRI pointings 1, 2, 3 and 6 were prime observations with NIRCam Imaging in parallel. MIRI pointings 5, 7, 8 and 9 were observed in parallel with NIRCam WFSS as prime.}
\end{deluxetable*}

\subsection{NIRCam WFSS Grism Survey}

We observe four of our NIRCam pointings (5, 7, 8, 9; Figure~\ref{fig:layout}), which have the greatest overlap with the NIRSpec pointings, with the NIRCam grism.  These data measure efficient redshifts of $z\!>$4 star-forming galaxies, whose emission lines (e.g.\ [OIII] $\lambda$5007) are expected to be strong. The NIRCam grism R$\sim$1500 data overlap the NIRSpec pointings, and enable us to calibrate NIRSpec slit losses at $\lambda =$3--4 $\mu$m, critical for absolute flux measurements. We used sources in the CANDELS EGS data to simulate the grism observations and optimize filter and central wavelength choices (minimizing sky background and spectral collisions). The F356W filter, which has a central wavelength close to the grism blaze wavelength (3.7 $\mu$m), combined with the Column (C) and Row (R) grism in both modules results in the greatest number of expected detected [O\,{\sc iii}] lines ($\sim$50 sources) from $z\!=$5.3--7.0 galaxies in our four pointings (the grisms in the B module are less sensitive by about 10\% and 15\% at the low and higher wavelengths, respectively). The grism data can also detect H$\alpha$ at $z$=3.8--5.1 and [O\,{\sc ii}] at $z$=7.5--9.7. These long-wavelength grism exposures were taken alongside the second set of short-wavelength F115W exposures, with MIRI imaging taken in parallel.  Each grism has a total integration time of 1245 sec, so sources detected in both observations will have a total time of 2490 sec.  We take shallow direct images both in and out of the field, to allow alignment with our deeper F356W observations described above, and identify the sources for all dispersed spectra.

\subsection{MIRI Survey}
The CEERS MIRI survey consists of eight pointings -- four as primary with NIRCam imaging in parallel (MIRI pointings 1, 2, 3 and 6), and four as parallel to the NIRCam grism observations (MIRI pointings 5, 7, 8 and 9).  In MIRI pointings 3, 6, 7 and 9, which overlap the NIRCam mosaic, we observe in F560W and F770W only, to probe the rest-frame optical emission (and thus stellar masses) of epoch-of-reionization galaxies (referred to as our ``MIRI-blue" pointings).  In the remaining pointings, which overlap existing {\it HST} and {\it Spitzer} imaging, we observe in the F770W, F1000W, F1280W, F1500W, F1800W and F2100W filters (referred to as our ``MIRI-red" pointings) to study the rest-mid-IR emission from galaxies (pointings 5 and 8, being in parallel to the NIRCam grism, have less exposure time available, thus we omit F770W and F2100W).   We use a 3-point dither pattern to enable MIRI self-calibration and adequate PSF sampling in parallel NIRCam images. 

All $\lambda >$ 8$\mu$m MIRI observations use the FAST readout mode, which the ETC predicts gives a $\sim$10\% higher SNR.  However, this mode has a high data volume, thus in the F560W and F770W, where the SNR gain is predicted to be less, we use the SLOW readout mode.   The exact exposures are a bit inhomogeneous due to the different parallel modes used.  For MIRI3 and 6, in parallel with NIRCam imaging, we use the SLOW readout mode with 41 groups/integration for each of the three dithers, which gives a total integration time of 2938 sec per observation.  F560W receives this observation once, while F770W receives it three times.  For MIRI7 and 9, to accommodate the NIRCam WFSS primary exposures, our base observation unit is SLOW readout with 12 groups/integration.  F560W receives 5 of these integrations during the grism R observing sequence (four during the primary NIRCam WFSS grism 4-point dither, and one during the direct image).  F770W receives 9 of these integrations (two during the grism R out-of-field image, four during the grism C 4-point dither, one during the grism C direct image, and 2 during the grism C out-of-field images).  The total integration times are 1433 sec in F560W, and 2580 sec in F770W. 

%% NIRSpec exposure table
\begin{deluxetable*}{lccccc}
\tablecaption{NIRSpec Exposure Times\label{tab:nirspec_exptimes}}
\tabletypesize{\small}
\tablehead{
\colhead{Instrument Pointing} & \colhead{PRISM/CLEAR} & \colhead{G140M/F100LP}
& \colhead{G235M/F170LP} & \colhead{G395M/F290LP} & \colhead{APA}
}
\startdata
NIRSpec4  & 3107.4 & 3107.4 & 3107.4 & 3107.4 & 89.32 \\
NIRSpec5  & 3107.4 & 3107.4 & 3107.4 & 3107.4 & 89.32 \\
NIRSpec7  & 3107.4 & 3107.4 & 3107.4 & 3107.4 & 89.32 \\
NIRSpec8  & 3107.4 & 3107.4 & 3107.4 & 3107.4 & 89.32 \\
NIRSpec9  & \nodata & 3107.4 & 3107.4 & 3107.4 & 89.32 \\
NIRSpec10 & \nodata & 3107.4 & 3107.4 & 3107.4 & 89.32 \\
NIRSpec11 & 6214.9 & \nodata & \nodata & \nodata & 41.43 \\
NIRSpec12 & 3107.4 & \nodata & \nodata & \nodata & 41.75 \\
DDT2750  & 18775.9 & \nodata & \nodata & \nodata & 352.25 \\
\enddata
\tablecomments{Exposure time (s) in each NIRSpec pointing and dispersion element. NIRSpec experienced an MSA short during the NIRSpec9 and 10 PRISM observations. Data exist for these pointings, but we do not include them in our data releases. The MSA configuration for NIRSpec11 was observed twice with a 1/3 shutter-width offset in the dispersion direction to test slit-loss corrections.}
\end{deluxetable*}

For MIRI pointings 1 and 2 we observe in F1000W, F1280W, and F1500W using 3 dithers of 100 groups x 2 integrations for 1673 s per filter. For F1800W we use 3 dithers of 40 groups x 5 integrations for 1698 s,  and for F2100W we use 3 dithers of 36 groups x 10 integrations and 3 dithers of 20 groups x 10 integrations for an exposure time of 4757 s.  Finally, we observe F770W in SLOW with 23 groups and one integration, for a total exposure time of 1648 s.  In MIRI pointings 5 and 8, in parallel to the NIRCam WFSS, we mimic the MIRI1/2 observing strategy within the constraints set by the prime WFSS exposures.  We end up with a base integration of FAST read with 112 groups, which is 311 sec.  We place four of these exposures (in parallel to the grism R 4-point dither pattern) in F1000W, which in parallel to the three grism R images (direct and out of field) we use F1280W.  We use F1800W in parallel to the grism C prime 4-point dither, and F1500W in parallel to the three grism C images.  The total integration times are 1243 s in F1000W and F1800W, and 922 s in F1280W and F1500W. We summarize the MIRI exposure times for each pointing and filter in Table~\ref{tab:miri_exptimes}.

\subsection{NIRSpec Survey}

The NIRSpec MSA observations address diverse goals: measuring redshifts of high-priority distant galaxies, characterizing ISM and AGN evolution, and testing \emph{JWST} observing strategies. To maximize the science gain and fully demonstrate NIRSpec as a survey workhorse, CEERS observes with both the R$\sim$100 prism, and the R$\sim$1000 ``medium resolution" (MR) gratings.  We opted for the latter over the highest resolution (R$\sim$2700) mode for improved sensitivity to continuum and faint line emission and (potentially) multiplexing, while still resolving lines that are blended at $R\approx 100$ (e.g., H$\alpha$+[N\,{\sc ii}]).
For the broadest redshifted emission-line coverage, we use all three R$\sim$1000 gratings.  Our pre-launch simulations predicted that for an $L^\ast$ $z$=6 galaxy, these data should detect $>$5 lines (including [O\,{\sc ii}], [Ne\,{\sc iii}], [O\,{\sc iii}], H$\beta$ and H$\alpha$).  We use 3-shutter slitlets with 3-point nodded exposures and ``midpoint'' shutter centering.  In each field we also observe with PRISM/CLEAR (R$\sim$100).  The six pointing centers were constrained by the layout of the parallel NIRCam mosaic.

We note that this 3-shutter nodding scheme moves objects diagonally across the NIRCam detector, subsampling pixels at three unique positions, which improves PSF reconstruction and size/morphology measurements, especially in the under-sampled F115W and F277W images.  However, this nodding scheme does not cover the NIRCam module gap.  Doing so would require additional exposures.  This would either reduce the integration per exposure, reducing the NIRSpec signal-to-noise, or would increase the program time beyond that suitable for an ERS program.  The gain in area by covering these gaps is a small fraction of the fiducial area, and thus does not affect the primary CEERS science goals, leading to the unique footprint shown in Figure~\ref{fig:fieldimage}.

The number of NIRSpec grating configurations is limited by the need to do NIRSpec grating/filter moves at the same time as the NIRCam filter moves, thus also requiring three observations.  In each observation, we observe in the PRISM, as well as the G140M/F100LP, G235M/F170LP, and G395M/F290LP medium-resolution gratings spanning 1-5\,$\mu$m, allowing the full range of NIRSpec science to be enabled.  The NIRSpec exposure times are set to be comparable to those for each of the NIRCam observations.  We choose three integrations of 14 groups each in STScI-recommended NRSIRS2 readout mode, giving a total exposure time of 3107 s.  
We summarize the NIRSpec exposure times for each pointing and dispersion element in Table~\ref{tab:nirspec_exptimes}.

During the NIRSpec prism observations of pointings 9 and 10, there was an electrical short in the MSA, which strongly illuminated the detector, contaminating the prism observations for both pointings.  We filed a Webb Operation Problem Reports (WOPR), and  these observations were approved for rescheduling (see \S 3.6 for discussion of original scheduling).  These two pointings were reobserved in February 2023 at different locations and V3PA (262.86 deg; additional NIRCam parallels were requested but not approved), and were renamed NIRSpec pointings 11 and 12.  We place NIRSpec pointing 11 near to NIRSpec pointing 10, to maximally overlap the NIRCam WFSS grism data.  In this pointing, in addition to the fiducial prism observation, we perform a second observation offsetting in the dispersion direction by 1/3 of a shutter width (originally planned for PRISM pointing 9, which was affected by the MSA short).  The combination of these two pointing 11 observations can be used to empirically calibrate wavelength-dependent differential slit losses versus object centering, using the NIRCam grism observations as an independent measure of the total line flux.  To maximize the science from pointing 12, we place it to the SW of the NIRCam mosaic to follow-up sources identified in the CEERS NIRCam imaging.

Potential NIRSpec targets were compiled and prioritized by members of the CEERS collaboration, particularly targeting $z >$ 4, where 1–5\,$\mu$m spectroscopy can detect key emission lines or continuum features to measure redshifts and spectral diagnostics. Higher weight was given to the highest redshift sources, and rare targets (such as quiescent or extreme emission line galaxy candidates). A larger sample of potential filler targets was also assembled, again prioritizing redshifts (mainly photometric) $z >$ 0.5 where H$\alpha$ and other strong lines are observable, and giving brighter filler objects higher weight. The final scheduling of CEERS (\S 3.6) resulted in some NIRSpec pointings overlapping CEERS NIRCam imaging from the first epoch, permitting observation of NIRCam-selected sources.  The remaining targets were selected primarily based on {\it HST} data.

\begin{figure*}[!t]
\epsscale{1.0}
\plotone{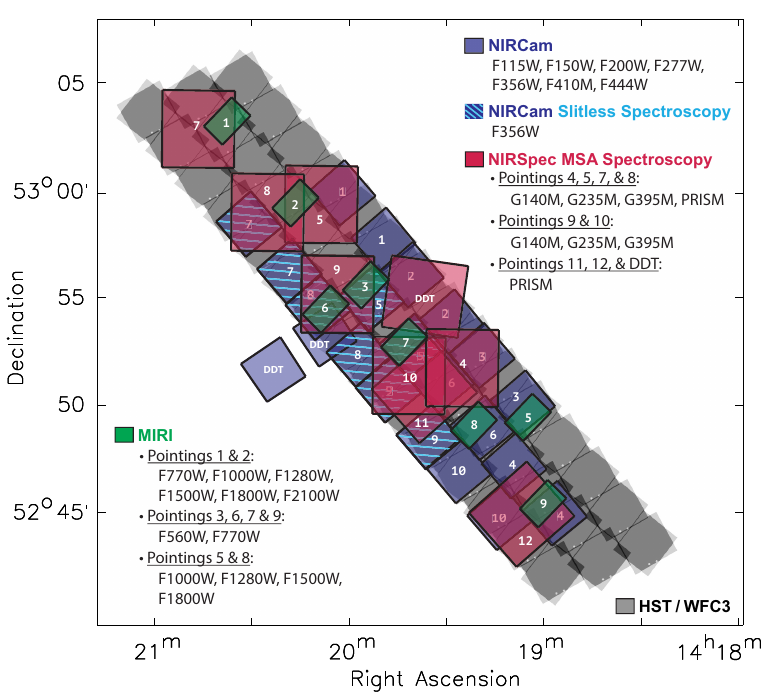}
\caption{The full observing layout of CEERS as executed, with the CANDELS {\it HST}/WFC3 footprint in the background.  The blue squares show the NIRCam survey, which consists of 10 pointings; as NIRCam has two modules, each pointing has a pair of identically numbered squares.  The four pointings with lighter-blue cross-hatching were also observed with the NIRCam WFSS grism elements, using the F356W filter.  The green rectangles show the eight MIRI pointings, while the larger red squares show the NIRSpec pointings.  This includes the six nominal NIRSpec pointings (4, 5, 7, 8, 9 and 10), as well as the prism-only pointings 11 and 12 (see \S 3.5), and the DDT followup deeper prism-only pointing \citep{arrabalharo23a}, with its parallel NIRCam pointing also shown.  The inset text summarizes the filters and spectral elements used in each pointing.}
\label{fig:layout}
\end{figure*}

\begin{figure*}[!t]
\centering
\epsscale{0.57}
\plotone{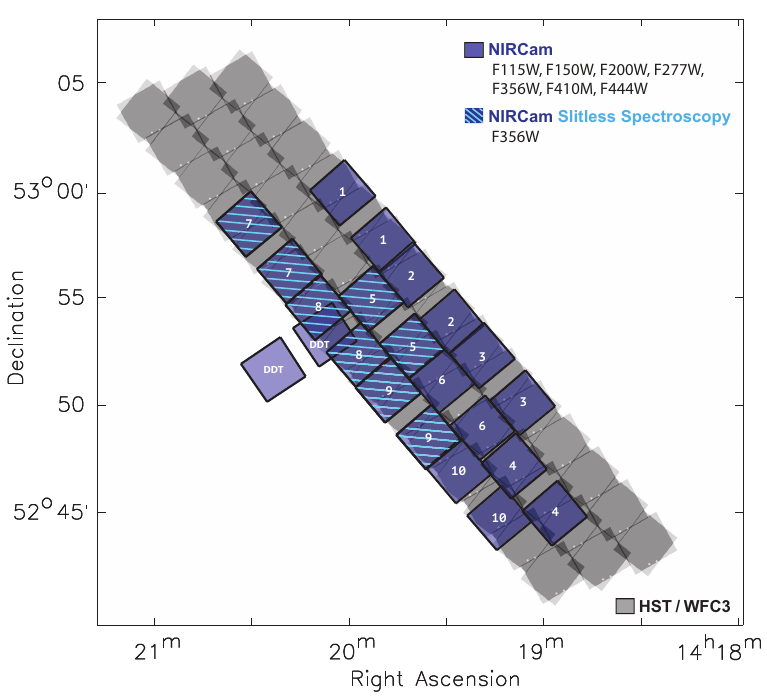}
\plotone{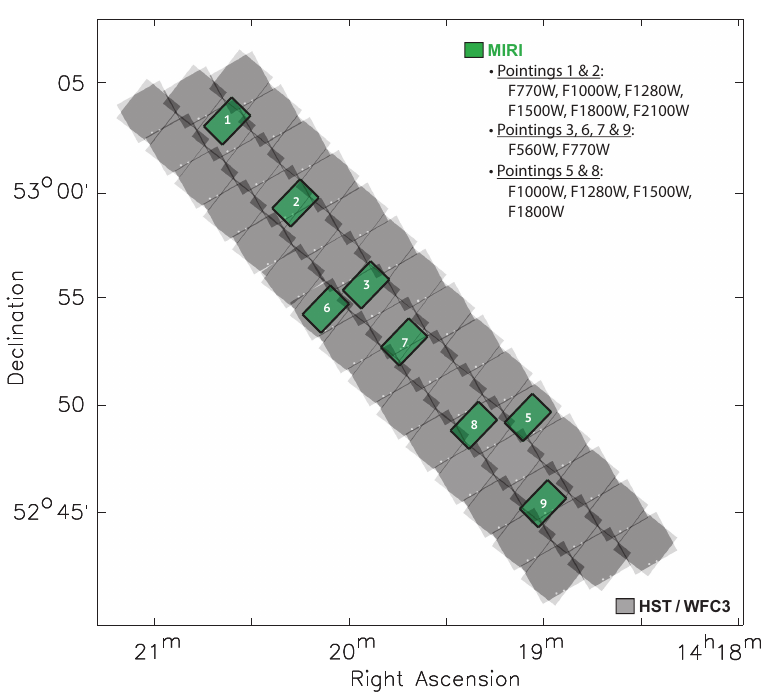}
\plotone{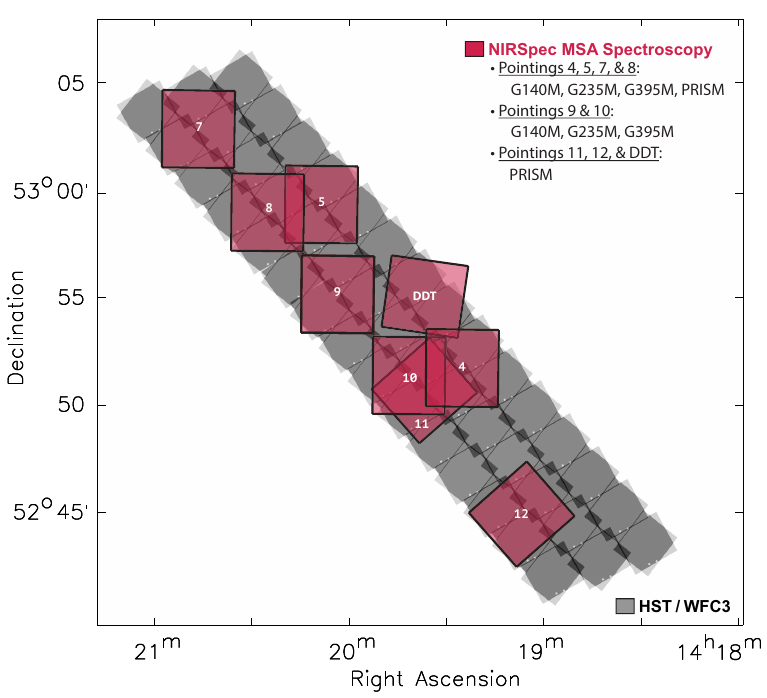}
\caption{Similar to Figure~\ref{fig:layout}, showing each instrument's layout separately.  Pointings for NIRCam are shown on the top-left, MIRI on the top-right, and NIRSpec on the bottom.  The background CANDELS {\it HST}/WFC3 footprint is again shown in gray.}
\label{fig:layout_inst}
\end{figure*}

The nominal NIRSpec field centers are set by the parallel locations relative to our NIRCam mosaic layout.  We further tweak the location of each NIRCam tile/NIRSpec MSA pointing within a few arcsec to optimize the yields of sources at the highest priorities.
Such optimization was performed with the MSA Planning Tool \citep[MPT][]{karakla14} allowing a deviation of the nominal pointing center in an 8 arcsec$^2$ box taking small steps of 0.05 arcsec ($\sim1/4$ of a shutter width). The resulting optimal centers were rechecked, applying tighter constraints to the few NIRSpec centers that disturbed the NIRCam parallel mosaic too much, causing undesired gaps or excessive overlapping in the imaging observations.
In the particular cases of NIRSpec11 and NIRSpec12, for which the centers were more flexible as they either had no NIRCam parallel observations or that was not intended to fit the rest of the NIRCam mosaic, we employed the eMPT tool \citep{bonaventura23} for an initial estimation of the nominal pointings and PAs. This choice was made based on the advantages eMPT presents at the time of efficiently exploring a large number of pointings and observing angles. From the nominal eMPT-derived locations for these three pointings, the fine tuning of the MSA configurations was further performed with the MPT in a similar way as described above for the other pointings.
In the end, 1089 different sources were observed in PRISM mode (1373 when including the sources in the two observations affected by the electrical short, for which limited information can be sometimes rescued), 313 with the MR gratings and 127 in both modes (200 considering the two defective prism observations). Note that the MR grating configurations at each pointing are identical for the three gratings employed, granting full 1-5 $\mu$m coverage for all the 313 sources observed in this mode.
A complete description of the NIRSpec target selection and MSA optimization is forthcoming in Arrabal Haro et al.\ (in prep).

\subsection{Observing Layout and Scheduling}
CEERS includes observations with three instruments over four instrumental modes, resulting in a mosaic of 10 NIRCam imaging pointings covering the majority ($\sim$90 arcmin$^2$) of the Extended Groth Strip HST legacy field.  Six of these pointings are in parallel to prime NIRSpec MSA spectroscopy, and four are in parallel to prime MIRI imaging.  Four of these pointings are also covered by NIRCam grism wide-field slitless spectroscopy (WFSS) with MIRI imaging in parallel, for a total of eight independent MIRI pointings.

%Placement of observations and position angle: 
We place our NIRCam mosaic along the bulk of the \textit{HST}/WFC3 region in the EGS field.  In order to place the NIRSpec parallels on the \textit{HST}-covered region (required for MSA pre-selection for an ERS program), we require a V3PA of $\sim$131 degrees for June and $\sim$311 degrees for December observing windows.  This position angle places our NIRCam mosaic parallel to the CANDELS/WFC3 coverage boundary, and maximizes the joint-overlap area between WFC3+NIRCam ($\sim$99\%) and WFC3+NIRSpec ($\sim$98\%).  The final specific position of the entire observing configuration at this PA is set by maximizing the coverage of our highest priority scientific sources (the seven pre-\textit{JWST} bright $z \gtrsim$ 9 candidate galaxies in this field) in our various instruments.  We note that the distribution of these rare sources places tight constraints on these observations; we arrived at our fiducial observation configuration by exploring a wide range of pointing centers and position angles.  With this optimal configuration, 5/7 candidates receive NIRCam imaging, 4/7 receive NIRSpec MSA spectroscopy, and two each receive MIRI imaging and NIRCam WFSS. Our fiducial PA is observable for $\sim$13 days in each of June and December.  We explored our tolerance for widening this PA constraint to ease scheduling, and conclude that unacceptable hits to the science achieved by our program begin to occur with PA changes of $\gtrsim$ 1~deg. 

At the time of proposal submission in 2017, the June window was optimal for an ERS program (allowing all observations to be taken within the first three months of science operations).  Subsequent launch delays also forced us to consider the December window, thus two complete versions of the full CEERS program were devised.  These complete ``alternate universe" plans are shown in Figure~\ref{fig:layout_alternate} in the Appendix.

The ultimate launch date of Dec 25, 2021 led to the entire CEERS program not being fully schedulable prior to the closure of the June window.  To meet the spirit of the ERS program, we requested, and were granted, the ability to split CEERS into two epochs, with a maximum of 25 hours being schedulable in June for early scheduling.  As it was unlikely the NIRSpec MSA would be fully commissioned by this time, we created a two-epoch version of CEERS, where four MIRI+NIRCam pointings would be executed at V3PA=131 deg in June (Epoch 1; pointings 1, 2, 3 and 6), and the remaining portion of the program would be executed at V3PA=311 deg in December (Epoch 2; pointings, 4, 5, 7, 8, 9, 10).  This precise 180 degree flip allows for our desired level of overlap between CEERS and existing datasets, and between different CEERS modes.  Additionally, the six-month separation allowed the December NIRSpec MSA plans to include {\it JWST}-discovered sources in the NIRCam imaging from Epoch 1, which led to a significant gain in scientific utility.
In fact, aiming to maximize the scientific return enabled by this two-epoch scheduling, pointing 4 was shifted all the way to the Southwest from its original position, preserving the continuity of the NIRCam mosaic while allowing a large overlap of the Epoch 2 NIRSpec4 pointing with the Epoch 1 NIRCam2, 3 and 6 observations (see Figures~\ref{fig:layout} and \ref{fig:layout_junedec}).

\begin{figure*}
\includegraphics[width=\textwidth]{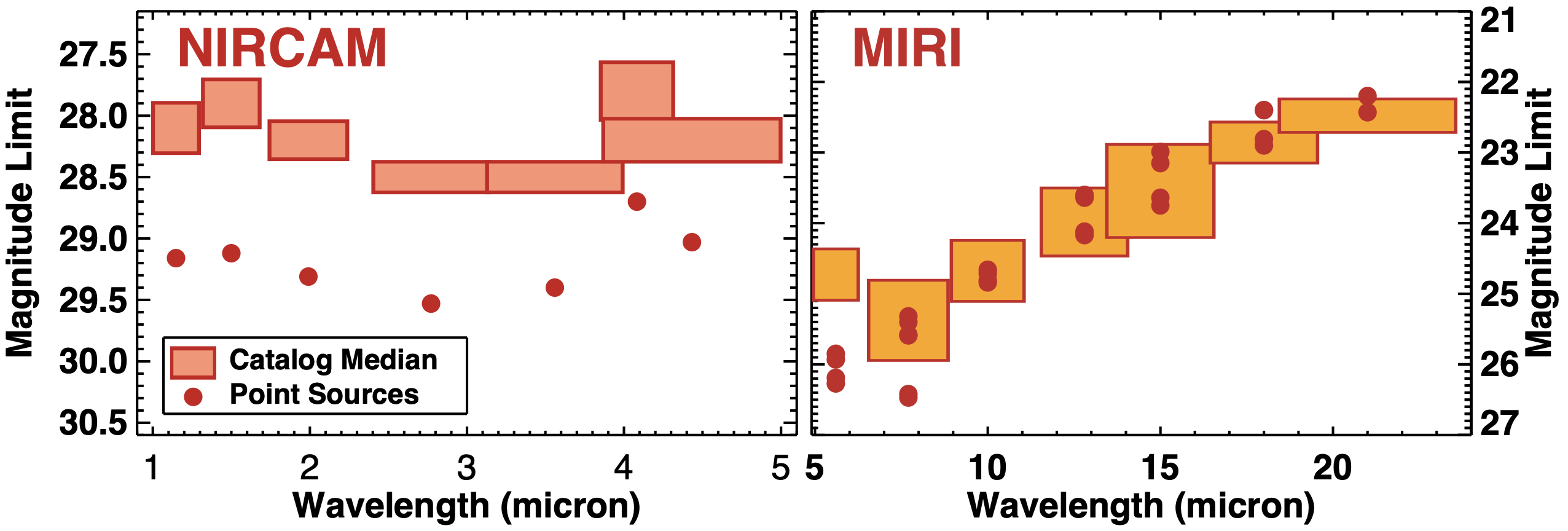}
\caption{Imaging depths as empirically measured from the released imaging from the CEERS team.  The point-source measurements were based on noise measured in randomly-placed 0.2\arcsec-diameter circular apertures, corrected to total based on the curve-of-growth as measured from the PSF for a given aperture size.  The catalog median is based on flux uncertainties from real sources in the respective photometric catalogs, where the vertical height denotes the 68\% spread, and the horizontal width denotes the filter FWHM.  All measurements are corrected to total for light falling outside a given aperture.  The MIRI catalog measurements converge to the PSF value at increasing wavelength due to the larger PSF.  Information on the photometric catalogs is available in I.\ Cox et al.\ (in prep) for NIRCam and \citet{yang23b} for MIRI.  The achieved NIRCam point-source depths are $\sim$0.3--0.5 mag deeper than pre-launch expectations.  The MIRI depths are also slightly deeper than expected for the bluer ($<$15$\mu$m); the reddest bands roughly match expectations.}
\label{fig:imagingdepth}
\end{figure*}

\begin{deluxetable}{ccc}
\vspace{2mm}
%\tabletypesize{\small}
\tablecaption{CEERS Imaging Data Quality Summary}
\tablewidth{\textwidth}
\tablehead{\multicolumn{1}{c}{Filter} & \multicolumn{1}{c}{Point Source} & \multicolumn{1}{c}{Median Catalog}\\
\multicolumn{1}{c}{$ $} & \multicolumn{1}{c}{5$\sigma$ Depth} & \multicolumn{1}{c}{5$\sigma$ Depth}}
\startdata
NIRCam F115W&29.2&28.1\\
NIRCam F150W&29.1&27.9\\
NIRCam F200W&29.3&28.2\\
NIRCam F277W&29.5&28.5\\
NIRCam F356W&29.4&28.5\\
NIRCam F410M&28.7&27.8\\
NIRCam F444W&29.0&28.2\\
\hline
MIRI F560W&25.9/26.2& 24.7\\
MIRI F770W&25.5/26.5&25.2\\
MIRI F1000W&24.8&24.6\\
MIRI F1280W&23.6/24.2& 23.9\\
MIRI F1500W&23.1/23.7&23.4\\
MIRI F1800W&22.4/22.9&22.8\\
MIRI F2100W&22.3&22.5
\enddata
\tablecomments{The NIRCam point-source depths are 5$\sigma$ limiting magnitudes, measured in d=0.2\arcs\ diameter circular apertures and corrected to total fluxes assuming a point-source.  The MIRI point-source depths are measured in circular apertures with diameters equal to the PSF FWHM, and corrected to total \citep{yang23b}.  As the MIRI pointings can have different exposure times (see Figure~\ref{fig:layout}), we provide two values where relevant.   The catalog median depths are the median flux errors in the respective photometric catalogs (which are all corrected to total); see Cox et al.\ (in prep) and \citet{yang23b} for more details.}
\label{tab:imagingdepth}
\vspace{-8mm}
\end{deluxetable}

\begin{deluxetable}{ccc}
\vspace{2mm}
%\tabletypesize{\small}
\tablecaption{CEERS Spectroscopic Data Quality Summary}
\tablewidth{\textwidth}
\tablehead{\multicolumn{1}{c}{Disperser/Filter} & \multicolumn{1}{c}{Emission Line} & \multicolumn{1}{c}{Continuum}\\
\multicolumn{1}{c}{$ $} & \multicolumn{1}{c}{5$\sigma$ Depth} & \multicolumn{1}{c}{5$\sigma$ Depth}\\
\multicolumn{1}{c}{$ $} & \multicolumn{1}{c}{(Unresolved)} & \multicolumn{1}{c}{(pix$^{-1}$)}\\
\multicolumn{1}{c}{$ $} & \multicolumn{1}{c}{(erg s$^{-1}$ cm$^{-2}$)} & \multicolumn{1}{c}{(AB mag)}}
\startdata
PRISM/CLEAR (1.5 $\mu$m)&7.3$\times$10$^{-18}$&26.5\\
PRISM/CLEAR (4.5 $\mu$m)&1.9$\times$10$^{-18}$&24.9\\
G140M/F100LP&2.2$\times$10$^{-18}$&24.1\\
G235M/F170LP&1.4$\times$10$^{-18}$&24.1\\
G395M/F290LP&1.1$\times$10$^{-18}$&23.9\\
NIRCam WFSS F356W&4.1$\times$10$^{-18}$&21.4
\enddata
\tablecomments{Depth of CEERS spectroscopic observations.  For the PRISM we list values at two representative wavelengths at the blue and red side; for the gratings/grism we list the median value across the full observation.  The emission-line depths come from simulations injecting mock unresolved emission lines to the data, and measuring the injected line flux which yields a 5$\sigma$ detection.  The continuum depths are per pixel; these generally become less sensitive at redder wavelengths as the wavelength width of a pixel decreases due to increased spectral resolution.}
\label{tab:specdepth}
\vspace{-8mm}
\end{deluxetable}

The final executed version of CEERS is shown in Figure~\ref{fig:layout}, which includes a single Director's Discretionary Time NIRSpec observation and corresponding NIRCam parallel executed in April 2023 (PI Arrabal Haro; PID DDT-2750).  In Figure~\ref{fig:layout_inst}, we show similar layout plots for each instrument individually, while in the Appendix, we show the June and December layouts separately in Figure~\ref{fig:layout_junedec}.

\section{CEERS Data Quality and Data Releases}

\subsection{Data Quality}

Here we analyze our internal reduction of all modes of the CEERS data, to provide characteristic depths for each type of observation.  Figure~\ref{fig:imagingdepth} shows our measured 5$\sigma$ depths for our NIRCam and MIRI imaging, with values tabulated in Table~\ref{tab:imagingdepth}.  In the left panel, for NIRCam, we show two different values.  The point-source depths are calculated using the empirically-measured noise directly from the images in 0.2\arcs-diameter apertures, correcting to total based on the fraction of flux from the PSF contained in this aperture in each filter.  The catalog median is based on real photometric sources from the CEERS NIRCam catalog (I. Cox et al, in prep), where we indicate the 68\% spread in the values as the height of the plotted rectangle; as most sources are resolved, these depths are naturally brighter (this catalog is PSF-matched to F444W; the catalog of \citealt{finkelstein24}, which was matched to F277W, has median uncertainties $\sim$0.3--0.5 mag deeper).  The achieved point-source depths are deeper by $\sim$0.3--0.5 mag compared to our pre-launch expectations from the CEERS proposal, where we stated that our imaging plan should yield 5$\sigma$ depths of $\sim$28.7 for unresolved sources.

A similar technique is used for MIRI in the right panel, where here we show the measurements done by \citet{yang23b}.  The point source measurements are done similarly to NIRCam, using randomly placed circular apertures in source-free regions; these were done separately for each MIRI pointing. The MIRI catalog measurements were done using Source Extractor for the two bluest bands (F560W and F770W), and TPHOT for the redder pointings (we direct the reader to \citealt{yang23b} for more details).  Here one can see the impact of the larger MIRI PSF, where at $\lambda \geq$ 10$\mu$m, the point-source and catalog depths are in agreement. 

\begin{figure*}
\includegraphics[width=\textwidth]{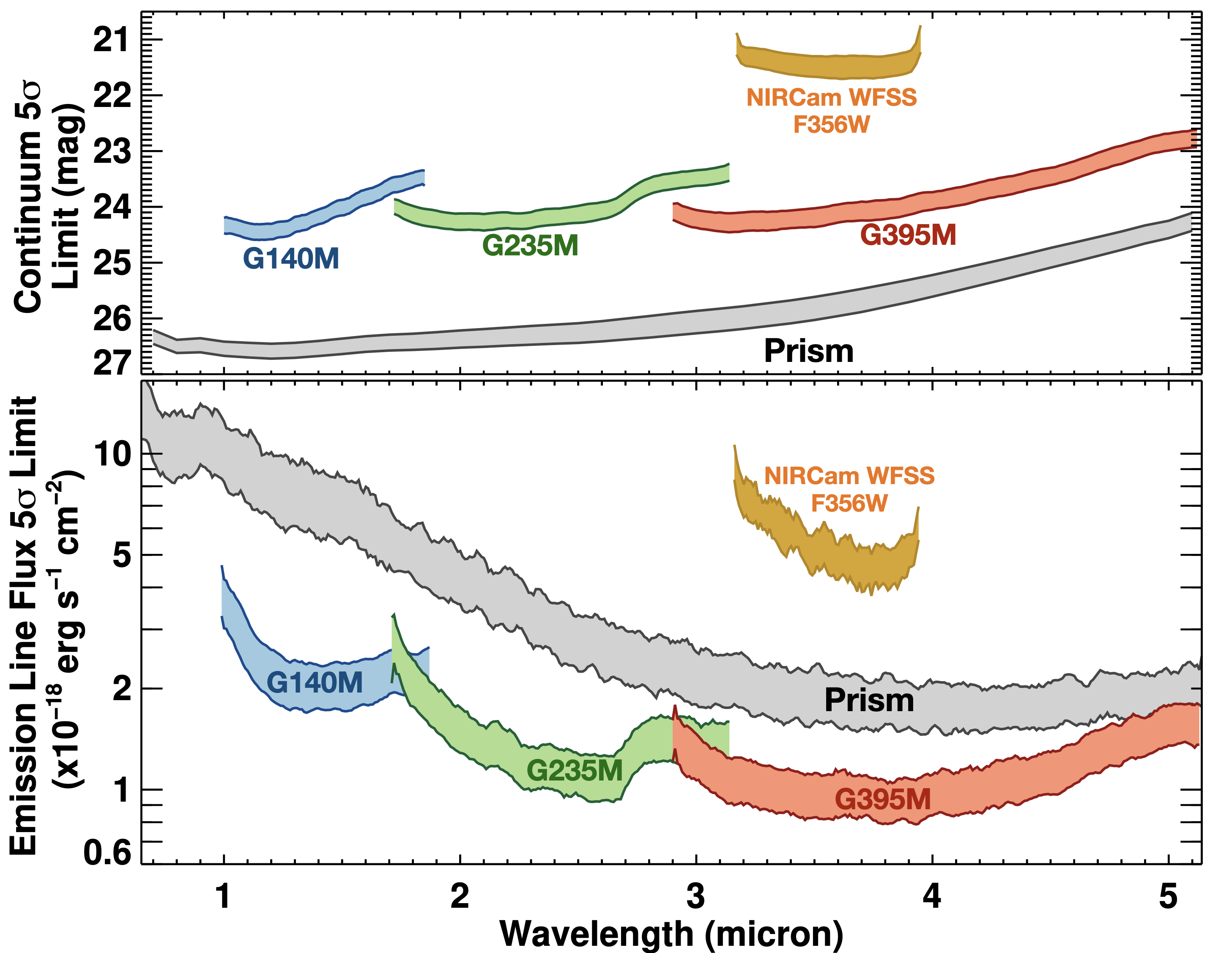}
\caption{Spectroscopic depths as empirically measured from the released spectroscopic data from the CEERS team.  Continuum measurements are based on the pixel-to-pixel noise from the reduced error arrays (and are thus shallower than what may be achieved when binning over multiple pixels).  The emission-line flux limits are measured empirically by placing mock emission lines in \emph{all} CEERS spectra, and recovering the median injected line flux where an emission line is recovered with a signal-to-noise of five.  The sensitivities of the various disperser elements are easily visible by the shapes of our sensitivity measurements.  Further details on these processes are described in \S 4.1.  }
\label{fig:specdepth}
\end{figure*}

Figure~\ref{fig:specdepth} shows our achieved spectroscopic depths, for continuum in the top panel, and emission lines in the bottom panel (tabulated in Table~\ref{tab:specdepth}).  For both NIRSpec and the NIRCam WFSS observations, we made use of the publicly released data products from DR0.7 (see \S\ref{sec:data_release_0.7}).  We estimated the continuum depth using the flux error arrays from each extracted spectrum, where for NIRSpec we included an additional correction of 1.7$\times$ into the error array, as empirically calibrated (see Appendix B in \citealt{arrabalharo23b}).  For emission line depths, we injected mock emission lines into the data for every spectrum, at each wavelength, with a range of emission line fluxes.  We then measured the resulting line flux with an MCMC algorithm, recording the recovered signal-to-noise.  At each wavelength we then calculated the injected emission-line-flux where the emission line was recovered at a signal-to-noise of 10, where half this value is the 5$\sigma$ limiting emission line flux.  This was done separately for each dispersion element (the three medium resolution NIRSpec gratings, the NIRSpec prism, and the NIRCam WFSS observations, including both Row and Column dispersers), where the shaded region in the figure is the 68\% in limiting flux values across all observations.  Similarly to the imaging, the achieved depths are somewhat better than expected in the proposal, where we claimed expected emission-line flux depths of $\sim$2 $\times$ 10$^{-18}$ erg s$^{-1}$ cm$^{-2}$ for the medium-resolution gratings.

\subsection{Data Releases}

To date the CEERS team has produced several data releases, from pre-launch simulated data products, through final releases for multiple datasets.  In this section, we briefly summarize the contents of each release.  All data releases are posted at the \href{https://ceers.github.io}{CEERS team's website} as well as on the \href{https://archive.stsci.edu/hlsp/ceers}{STScI MAST archive}.

\subsubsection{Simulated Data Release v1 (SDR1)}
In our first data release we provided simulated raw CEERS data for one pointing of NIRCam imaging and six pointings of MIRI imaging. We provided readme files for each instrument mode, and Jupyter notebooks showing how to reduce the raw data through the JWST Calibration Pipeline. We simulated NIRCam images for one CEERS pointing (CEERS 5) using MIRAGE\footnote{mirage-data-simulator.readthedocs.io} version 2.1.0, with input sources taken from a mock catalog created using the Santa Cruz semi-analytic model for galaxy evolution \citep{yung22}. The images were simulated using pointing and XML files based on the CEERS APT (with modifications required for MIRAGE to simulate the custom primary-parallel dither patterns planned for CEERS observations). We simulated MIRI images for all six CEERS pointings using MIRISIM version 2.4.0, with input sources taken from the same SAM mock catalog.  We presented these data products as part of \href{https://www.stsci.edu/jwst/science-execution/jwebbinars}{JWebbinar} 13: CEERS NIRCam and MIRI Imaging.

\subsubsection{HST Data Release v1 (HDR1)}
In this data release we provide updated {\it HST} mosaics in the EGS field.  These mosaics were produced following similar procedures to those described in \citet{koekemoer11}, in particular incorporating improvements in calibration and astrometry beyond those available from the default {\it HST} archive pipeline products. The absolute astrometric frame for these mosaics is Gaia-EDR3, where we used catalogs from the DESI Legacy Survey imaging \citep{dey19}, itself tied to Gaia, as an intermediate astrometric reference.  All the mosaics are pixel aligned to each other on this astrometric frame to a precision of a few milliarcseconds. The mosaics combine data in 6 filters (ACS/WFC F606W, F814W and WFC3/IR F105W, F125W, F140W, F160W) from a total of 1,767 exposures, obtained from eight different \textit{HST} programs (10134, 12063, 12099, 12167, 12177, 12547, 13063, 13792), with all the mosaics at a pixel scale of 30 milliarcseconds/pixel. 

\subsubsection{Simulated Data Release v2 (SDR2)}
In this second simulated data release, we provided simulated NIRSpec MSA observations for one CEERS pointing, including fully reduced and calibrated spectra in all CEERS filter/grating configurations as well as Jupyter notebooks showing how to reduce the data through the JWST Calibration Pipeline.  We simulated NIRSpec MSA data for one CEERS NIRSpec pointing using the NIRSpec Instrument Performance Simulator \citep[IPS;][]{piqueras10}, with input sources taken from the CANDELS EGS photometric catalog \citep{stefanon17}. The input spectra are simulated as continuum SED models with emission lines using the MAPPINGS library \citep{allen08}, with simulation scenes created from the CEERS MSA configurations. The IPS turns these scenes into simulated count-rate maps, which are the input for Stage 2 of the JWST Calibration Pipeline.

\subsubsection{Simulated Data Release v3 (SDR3)}
In this final simulated data release, we provided simulated NIRCam WFSS observations for four CEERS pointings. We provided the input models, raw data, calibrated count rate maps, and extracted spectra. We also provided updated NIRCam imaging in one CEERS pointing. This NIRCam imaging update includes improved input photometry, additional depth in two filters, and pixel-aligned mosaics in all filters.  The simulated NIRCam WFSS observations and the associated imaging for all four CEERS pointings (CEERS 5-8) were created using MIRAGE version 2.2.1, with input sources taken from the SAM mock catalog. NIRCam WFSS exposures in each grism are taken with one LWC direct image followed by two LWC out-of-field images to identify the sources associated with the spectra. During all CEERS WFSS observations and LWC imaging, the SWC observes with F115W to add depth to the NIRCam imaging in these fields.

\subsubsection{CEERS Data Release v0.5}
In November 2022, we provided our team's reductions of the first epoch (June 2022) of CEERS observations, as well as documentation and the scripts and Python code we used to reduce the images. 
We reduced the images through version 1.7.2 of the JWST Calibration Pipeline \citep{bushouse24} and Calibration Reference Data System (CRDS) pmaps 0989 (NIRCam) and 0970 (MIRI), with custom Python scripts developed to handle additional corrections such as ``snowballs'', ``wisps'', $1/f$ noise (for NIRCam), and outlier detection and background subtraction (for MIRI). See \citet{bagley23b} and \citet{yang23b} for details on this initial reduction. 

\subsubsection{CEERS Data Release v0.6}
In May 2023, we provided our team's reductions of the second set of CEERS MIRI and NIRCam imaging, observed as part of CEERS Epoch 2 in December 2022. This data release included CEERS NIRCam pointings 4, 5, 7, 8, 9 and 10, obtained in parallel to prime NIRSpec MSA observations. We also included the SW F115W and direct imaging in F356W obtained as part of the NIRCam WFSS observations in pointings 5, 7, 8 and 9. This DR0.6 NIRCam reduction is almost identical to that of the DR0.5 release, where the difference is an updated version of the Calibration Pipeline (1.8.5) and CRDS pmap (1023). The DR0.6 MIRI release included all eight MIRI pointings from both CEERS epochs, uniformly reduced with Calibration Pipeline version 1.9.3 and CRDS pmap 1039.

\subsubsection{CEERS Data Release v0.7}
\label{sec:data_release_0.7}
In October 2023, we provided our team's reductions of the NIRSpec MSA and NIRCam slitless grism spectra, observed as part of CEERS Epoch 2 in December 2022 and Epoch 3 in February 2023.  This data release included CEERS NIRSpec pointings 4, 5, 7, 8, 9 and 10, obtained as prime observations with NIRCam imaging parallels. The PRISM observations in pointings 9 and 10 were affected by an MSA short in December and were repeated at new positions in February (comprising Epoch 3). These two new PRISM pointings are 11 and 12.  We reduced these spectra using custom procedures that improve upon the default pipeline, including custom aperture extractions and masking of detector artifacts and other contaminants. These improvements will be described in detail in Arrabal Haro et al.\ (in prep).

\begin{figure*}[!t]
\centering
\epsscale{1.15}
\plotone{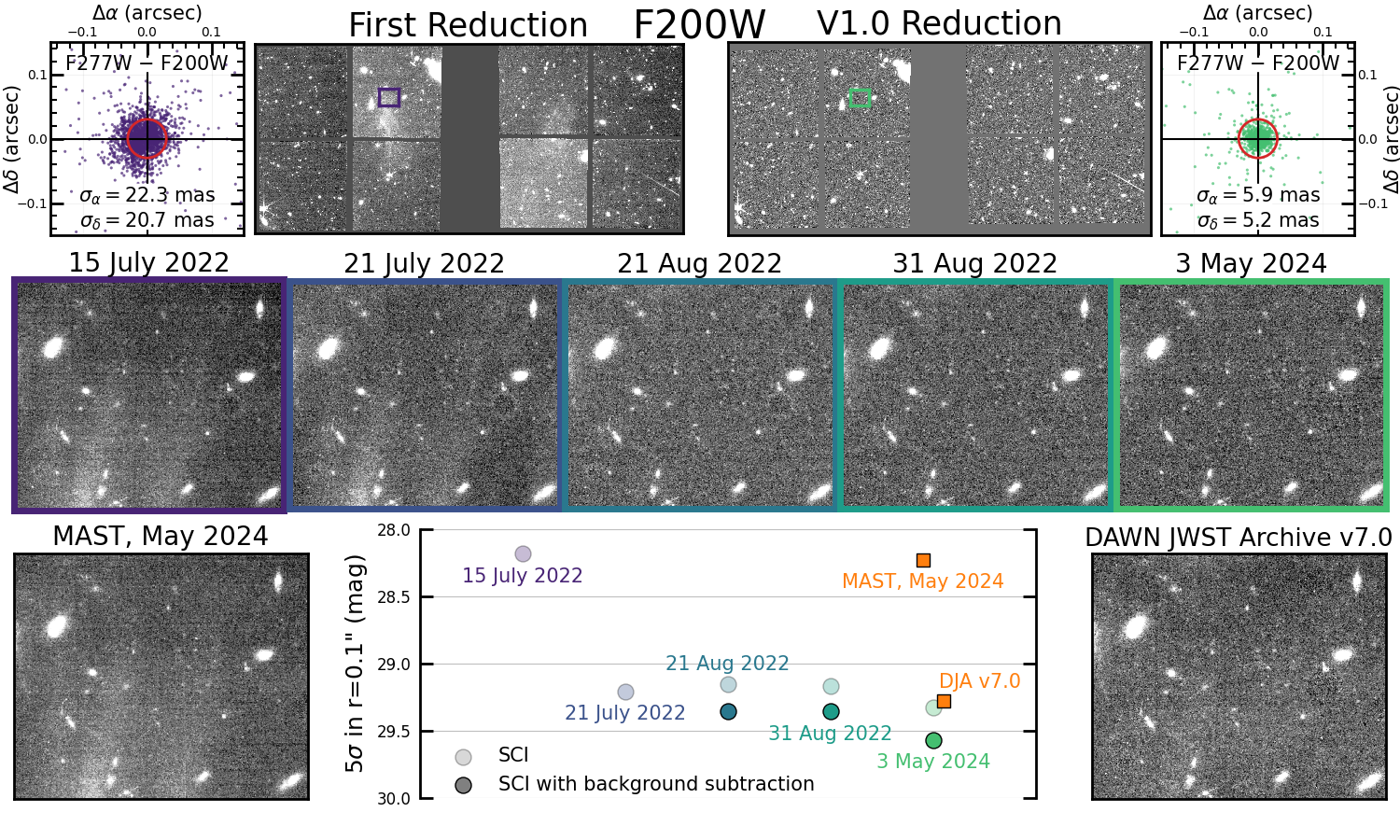}
\caption{
A schematic representation of the progression of CEERS NIRCam data reduction, using F200W as an example. Images in the top row contrast mosaics for a CEERS pointing produced by our first reduction in July 2022, and our v1.0 reduction in May 2024, including an astrometric improvement (top outer panels) showing the RMS of the scatter between F277W and F200W reducing from $\gtrsim20$mas to $\sim$5--6 mas. The middle row shows stamps of a zoom-in region through five reduction versions, where iterative improvements include corrections for snowballs, wisps and $1/f$ noise as well as a 2D background subtraction. The same pointing downloaded from MAST (lower left) and the DAWN JWST Archive (v7.0, lower right) are included for comparison. The $5\sigma$ limiting magnitudes measured in $r=0\farcs1$ apertures on each reduction are plotted in the bottom middle panel, with an improvement of $>$1.4 mag from the first reduction. The CEERS reduction is a significant improvement over that available in MAST, and is $\sim$0.3 mag deeper than the DJA mosaic due primarily to our 2D background subtraction.
\label{fig:nircamredux}}
\end{figure*}

\subsubsection{CEERS Data Release v1.0}
The forthcoming CEERS v1.0 data release will include our final reduced version of our imaging and spectroscopic data products, as well as a NIRCam photometric catalog (I.\ Cox et al, in prep).  The spectroscopic datasets will be described in forthcoming papers by Arrabal Haro (in prep) and N.\ Pirzkal (in prep).  As NIRCam \citep{bagley23b} and MIRI \citep{yang23b} data papers are already published, we share here in the appendix an updated description of our imaging reduction processes, for NIRCam in Appendix B, and MIRI in Appendix C.  In Figure~\ref{fig:nircamredux} we show the ``evolution" of the quality of our internal NIRCam reductions, from the first days in Summer 2022, through our final v1.0 reduction, where the increase in depth in F200W was nearly 1.5 magnitudes.  We also compare to the latest reductions from the MAST pipeline, and the DAWN JWST Archive (DJA), finding that our custom reduction performs significantly better than the MAST archive, with more modest improvements when compared to the DJA.

\begin{figure*}[!t]
\centering
\epsscale{1.15}
\plotone{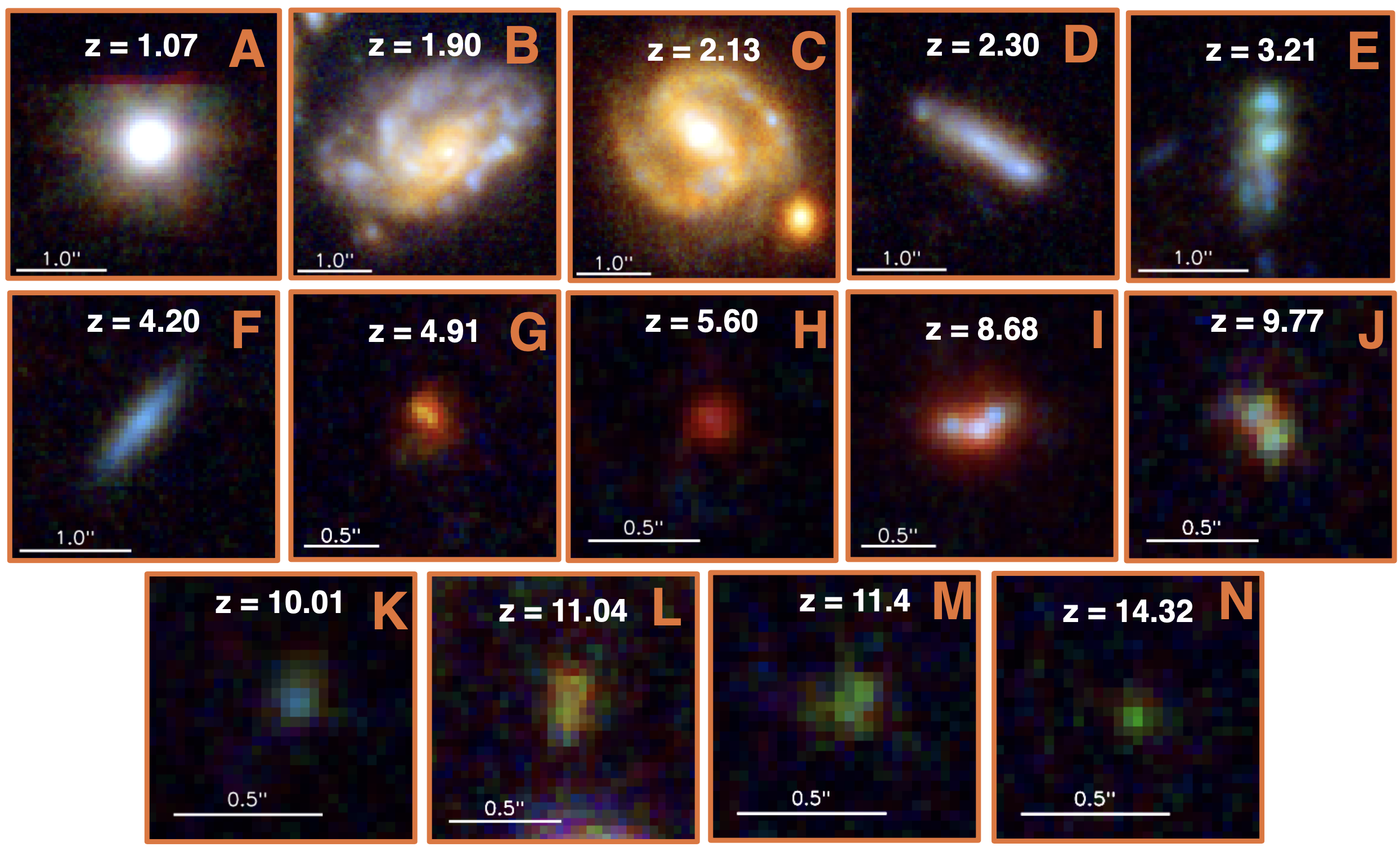}
\caption{NIRCam imaging highlights from CEERS.  The objects shown are: $a$) A $z=1.07$ galaxy (\S \ref{gal-property}) which exhibits strong spectroscopic signatures of the thermally pulsing asymptotic giant branch (TP-AGB) 
stars \citep{lu24}.  $b$) A $z =$ 1.957 galaxy from \cite{lebail23}, where CEERS imaging reveals a red bulge surrounded by $HST$-observed spiral arms. $c$) A barred spiral galaxy at $z =$ 2.13 (\S 5.3), which was the first known $z >$ 2 barred spiral \citep{guo23}.  $d$) A prolate galaxy candidate at $z \sim$ 2.3, part of a large sample of such objects from \citet{pandya24}, showing galaxies at $z \sim$ 1--8 to be preferentially prolate. $e$) and $f$) A $z \sim$ 3.2 irregular galaxy and a $z \sim$ 4.2 disk-galaxy (respectively), showing both the diversity of morphologies in the early universe, as well as the surprising prevalence of disk galaxies at high redshift \citep{kartaltepe23}.  $g$) A $z = 4.9$ galaxy originally thought to be at $z \sim$ 16 \citep[e.g.][]{donnan23a}, shown via spectroscopy to be at this much lower redshift, with a unique combination of strong emission lines causing the redshift overestimate \citep{arrabalharo23a}.  $h$) The first confirmed high-redshift broad-line AGN from {\it JWST}, as well as the first known ``little red dot", from \citet{kocevski23b}. $i$) A galaxy at $z \sim$ 8.7 which is believed to host an AGN, the highest-redshift AGN at the time of publication \citep{larson23a}.  $j$), $k$) and $l$) Spectroscopically confirmed very high-redshift galaxies at $z =$ 9.77, $z =$ 10.01 \citep{arrabalharo23b}, and $z =$ 11.04 \citep{harikane24}.  $m$) Maisie's Galaxy, the first early \textit{JWST} photometric candidate (c.\ July 2022) to be spectroscopically confirmed
\citep{arrabalharo23a}, at $z =$ 11.416. $n$) A $z =$ 14.3 candidate galaxy \citep{finkelstein24}.  Objects $a$, $h$, $j$, and $m$ also have spectra shown in Figure~\ref{fig:specfig}.}
\label{fig:imagingfig}
\end{figure*}

\section{Science Highlights}

\begin{figure*}[!t]
\centering
\epsscale{1.1}
\plotone{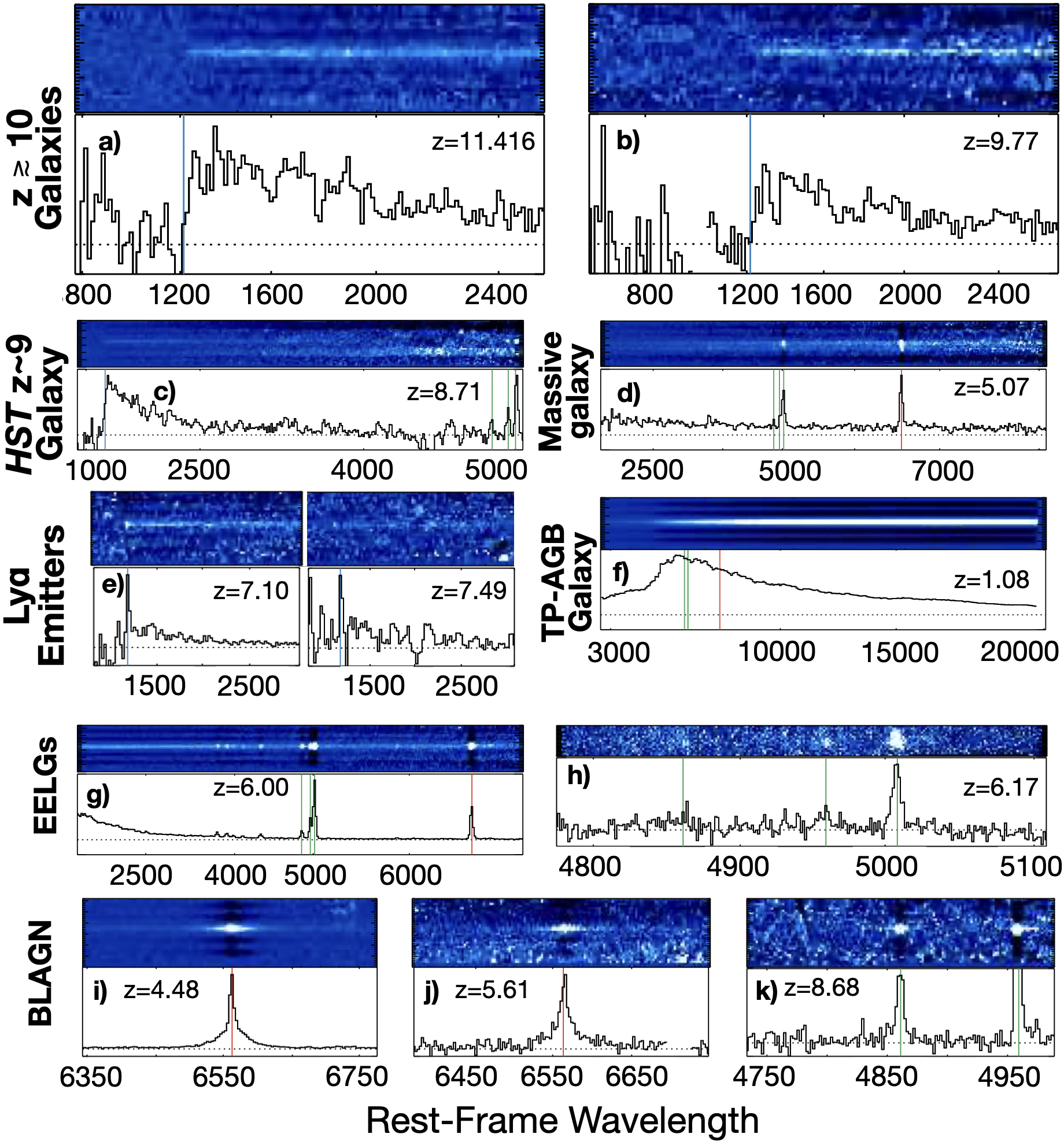}
\caption{A compilation of spectroscopic discoveries from CEERS.  The top row shows prism spectra for two spectroscopically confirmed galaxies; panel $a$) shows ``Maisie's Galaxy", the first high-redshift discovery from CEERS, and the first early \textit{JWST} photometric candidate to be confirmed \citep{arrabalharo23a}.  The right panel shows a $z \sim$ 10 candidate from \citet{finkelstein23} confirmed to be at $z =$ 9.77 \citep{arrabalharo23b}.  The second row shows $c$) a $z \sim$ 9 $HST$-selected candidate from \citet{finkelstein22}, used as an example object in the CEERS proposal, confirmed to be at $z =$ 8.71 \citep[][Larson et al., in prep]{tang23}, and $d$) a massive galaxy from \citet{chworowsky24}. 
The third row shows $e$) two Ly$\alpha$ emitters from \citet[][left]{napolitano24} and \citet[][right]{chen24}, and $f$) A galaxy with significant TP-AGB star emission (\S \ref{gal-property}) at $z =$ 1.08 \citep{lu24}.  The fourth row shows two extreme emission-line galaxies, one from NIRSpec ($g$) from \citet{davis24}, and one from NIRCam grism ($h$) from \citet{backhaus24}.  The final row shows G395M spectra for three discovered broad-line AGN: $i$) at $z =$ 4.48 from \citet{harikane23b}; $j$) at $z =$ 5.61 from \citet{kocevski23b}, and $k$) at $z =$ 8.68 from \citet{larson23a}; the first two are based on strong broad H$\alpha$, while the latter is based on weaker (2.5$\sigma$ significance) broad H$\beta$.}
\label{fig:specfig}
\end{figure*}

In this section we highlight some of the key early science results from CEERS, including both results from the CEERS team, as well as a sampling of early results from the community.  We summarize these discoveries in Figures~\ref{fig:imagingfig} and \ref{fig:specfig}, which show imaging and spectroscopic highlights, respectively.  Several of these results come from a set of "Key Papers" written by the CEERS team; we provide a full listing of these key papers in Appendix A. We finish this section with an analysis of publication statistics from all papers published or submitted by Dec 2024 using CEERS data.

\subsection{Epoch of Reionization}

The primary goal of CEERS was to characterize the $z >$ 10 universe, and explore what could be learned to higher redshifts.  In the days after CEERS data were released, two papers were quickly written on exciting very high-redshift universe objects (coming just a few days after the very first {\it JWST} very high-redshift discoveries from \citealt{castellano22} and \citealt{naidu22} from the GLASS ERS program data).  From the CEERS team, in \citet{finkelstein22c} we showed the discovery of Maisie's Galaxy, a candidate $z \sim$ 12 galaxy in CEERS Epoch 1 imaging.  This discovery was surprising, as pre-launch theoretical simulations did not predict an object at this luminosity and redshift would be detectable in the modest CEERS survey parameter space.  Contemporaneously \citet{donnan23a} showed the discovery of an even more distant object, a candidate $z \sim$ 16 galaxy, nearly one magnitude brighter than Maisie's Galaxy.  These two objects were so exciting that a Director's Discretionary Time program was awarded (PI Arrabal Haro; JWST-DDT 2750) to obtain rapid spectroscopic redshifts.  While Maisie's Galaxy was confirmed to be at $z_{\mathrm{spec}}=$11.416, the $z \sim$ 16 candidate was shown to be a $z =$ 4.9 interloper, with a diabolical pair of extremely strong [\ion{O}{3}] and H$\alpha$ emission lines which contaminated four of the NIRCam bands, mimicking the photometric signature of a $z \sim$ 16 Ly$\alpha$ break \citep{arrabalharo23a}, a scenario previously discussed by the community as a possible explanation for such an exotic object \citep[e.g.,][]{zavala22,naidu22b}.

Within a few months, larger samples of galaxies were discovered in the CEERS field, allowing the construction of rest-frame UV luminosity functions.  These could be used to settle the long-standing debate about whether the abundance of galaxies at $z >$ 9 continued to follow the smooth decline \citep{coe13,finkelstein16,mcleod16} observed from $z =$ 4--8, or followed an accelerated decline \citep{oesch16,bouwens19}.
In a pair of papers by the CEERS team \citep{finkelstein23,finkelstein24}, we showed that Maisie's Galaxy had many friends, ultimately publishing a sample of 88 galaxy candidates at 8.5 $< z <$ 14.  UV luminosity functions were constructed at $z \sim$ 9, 11 and 14, and rather than an accelerated decline, or even smooth decline, we found that the evolution of the number density of bright ($M_{UV}=-$20) galaxies was shallowing with increasing redshift at $z >$ 9.  

Similar results were also seen in studies by several other authors using these data \citep[e.g.][]{mcleod23,harikane23,adams23}, as well as in other fields \citep[e.g.][]{franco24,casey23b,castellano23,bouwens23b,hainline23,robertson24,donnan24}.  While this surprising result could indicate significant contamination in the samples, 13 of these galaxies were spectroscopically confirmed by CEERS or the followup DDT program \citep{fujimoto23,larson23a,arrabalharo23a,arrabalharo23b,harikane24}, with only one interloper (discussed above).  This implies that explaining the observations may require changes to the dominant physical processes regulating star formation, with the data consistent with predictions from models that naturally have enhanced star-formation efficiency and/or stochasticity. 

While more data is needed to get to the bottom of what physical processes are responsible, measurements of stellar masses at slightly lower redshifts also imply changing physical processes.  An early paper by \citet{labbe23} searched for massive high-redshift galaxies, uncovering a population of 13 extremely massive galaxies.  While originally these results were interpreted as uncovering a population of galaxies ``too massive" for $\Lambda$CDM cosmology \citep[e.g.][]{mbk23}, further analysis showed that many of these galaxies likely had rest-frame optical emission dominated by AGN emission (discussed further in \S 5.2).  In \citet{chworowsky24}, the CEERS team explored the evolution of the abundance of massive (log M/M\sol\ $>$ 10) galaxies from $z \sim$ 1--7.  They found that when likely AGN are excluded, there was no tension with $\Lambda$CDM expectations.  They did find that while from $z =$ 1--4 the evolution of this abundance was consistent with that expected from the evolution of the dark matter halo mass function combined with a fixed conversion of baryons into stars (of $\sim$14\%), from $z \sim$ 4--7 this evolution of the abundance shallows.  This observation could indicate either an inferred increase in the stellar baryon fraction to roughly double the $z =$ 4 value by $z =$ 7, or a reduction in the typical mass-to-light ratio (again by a factor of $\sim$2) over this same interval.  Either of these scenarios, extrapolated to higher redshift, could be consistent with the high abundance of UV luminous systems discovered at $z >$ 10.

Spectroscopic redshifts were just the tip of the iceberg for science investigations into the CEERS spectroscopic dataset.  In the reionization epoch, the detectability of Ly$\alpha$ emission was of immense interest due to its sensitivity to a neutral IGM.  \citet{napolitano24} explored the full CEERS prism dataset to discover whether any evolution in Ly$\alpha$ emission was detectable.  Surprisingly (given the reduced NIRSpec sensitivity at the blue end), they found a sample of 65 Ly$\alpha$ lines at 4 $< z <$ 8.5.  They calculated the fraction of galaxies that exhibit detectable Ly$\alpha$ emission, and found little evolution from $z =$ 5 to 7 in the CEERS field, though the $z =$ 7 value is lower when spectroscopy from JADES in the GOODS-S field is included, hinting at the impact of large ionized bubbles in the CEERS field.  The presence of such bubbles is further explored by \citet{chen24}, who also studied Ly$\alpha$ emission in the CEERS dataset, finding evidence for overlapping ionized bubbles along the line of sight from $z \sim$ 7.1 to 7.8 in this field (see also \citealt{tang23,nakane24}).  \citet{napolitano24} noted that the Ly$\alpha$ fraction at $z \sim$ 5 was lower than values from previous ground-based observations, which could indicate that a significant fraction of the emergent Ly$\alpha$ emission from high-redshift galaxies is extended, thus falling outside the small NIRSpec MSA slits (0.2\arcs\ $\times$ 0.4\arcs, compared to typical ground-based seeing and/or slit width of $\sim$1\arcs).

The CEERS survey further allows the detailed analysis of the complexity and diversity of Ly$\alpha$ escape during the epoch of reionization. Analyzing rest-frame UV to optical spectra of bright Ly$\alpha$ emitters, \citet{jung24} reveal highly ionized and metal-poor ISM in these galaxies. The study highlights significant variations in Ly$\alpha$ flux, velocity offset, and spatial extension, indicating diverse Ly$\alpha$ escape mechanisms across different Ly$\alpha$ emitters. To facilitate the escape of Ly$\alpha$, the source with the highest ionization may ionize its own bubble, while others require additional ionizing sources to create sizeable ionized bubbles. These findings suggest varying scenarios for the creation of ionized bubbles during reionization, underscoring the need for high-spectral-resolution data to further understand these processes in greater detail.

CEERS spectra were also useful in uncovering the physical properties of high-redshift galaxies, moving beyond the rest-UV for the first time.  %\textcolor{red}{Text on metallicities from Seiji, include Nakajima [though mentioned in Sec 5.4 briefly too], Isobe, others?}
In \cite{fujimoto23}, strong rest-optical emission lines of \OIII\ and H$\beta$ are successfully detected among $z\simeq$ 8--9 galaxies found in CEERS, characterizing those NIRCam-discovered early galaxies generally with high \OIII+H$\beta$ equivalent width (EW$\simeq$1100${\rm \AA}$) and elevated ionizing photon production efficiency ($\log(\xi_{\rm ion}/{\rm Hz~erg^{-1}})\simeq25.8$), which highlights their key contributions to the cosmic reionization (similar results have been found in other fields, e.g., \citealt{simmonds24,zavala24}).  The empirical calibrations of \OIII/H$\beta$ suggest that these galaxies have higher SFRs and lower gas-phase metallicity than those with similar stellar mass at $z\sim$ 2--6, which is consistent with simulation predictions and later statistical works including lensing cluster fields by \cite{nakajima23}. 
Using the same sample as \cite{nakajima23} and securely determining the line spread function of the medium grating data of NIRSpec, \cite{isobe23} also explore the electron density measurement with \OII$\lambda\lambda3726,3729$ doublets. The authors find an increasing trend of $n_{\rm e}$, observed so far from $z\sim0$ to 2, continues to $z\sim9$ likely due to the compact morphology of high-$z$ galaxies. Finally, \citet{mascia24} used the spectroscopic properties of high-redshift galaxies observed from CEERS to predict their ionizing photon escape fractions, using indirect diagnostics calibrated at lower redshift.  They found that the typical implied escape fraction was $\sim$13\%, implying that observable galaxies contribute significantly to the reionization process.

\subsection{Active Galactic Nuclei}

One of the more surprising results to come from the CEERS observations is the detection of numerous faint, broad-line AGN at $z>5$.  The first of these objects was identified photometrically as a candidate low-luminosity quasar by \cite{onoue23} and spectroscopically confirmed by \cite{kocevski23}.  Using the CEERS NIRSpec G395M/F290LP spectroscopy, \cite{kocevski23} found broad H$\alpha$ emission in two sources at $z\sim5$ and measured their BH masses to be $10^{6-7}$ M$_{\odot}$, making these sources the least massive BHs known in the early universe.  Soon thereafter, \cite{larson23b} reported broad H$\beta$ emission in a galaxy at $z=8.679$, making it, at the time, the most distant AGN ever identified.  A subsequent survey of broad-line AGN in CEERS by \cite{harikane23} reported the number density of these sources to be 1--2 dex higher than that of bright quasars identified by ground-based surveys at similar redshifts (see also \citealt{matthee23, maiolino23b}).

One of the sources reported in \cite{kocevski23} was initially identified as a candidate massive galaxy at $z=8.13$ by \cite{labbe23} and instead found to be a heavily reddened, broad-line AGN at $z=5.624$.  The source features a steep red continuum in the rest-frame optical and relatively blue colors in the rest-frame UV.  Sources with this ``v-shaped", red-plus-blue spectral energy distribution (SED) have come to be known as “little red dots” (LRDs) in the literature \citep{matthee23}.  

These heavily obscured sources have been studied extensively with the CEERS dataset.  \cite{Barro23} performed the first photometric selection of these sources using color cuts to pick out their unique SED shape (see also \citealt{labbe23b, kokorev24}).  More recently, \cite{Kocevski24} selected a sample of LRDs in CEERS (and other public fields) based on their rest-frame optical and UV continuum slopes and showed that the population appears to emerge in large numbers at $z\sim5$, spanning $\sim 1$ Gyr years of cosmic history, from rise to fall.  Using public data from the RUBIES survey (GO-4233; PI: A. de Graaff, \citealt{degraaff24}), \cite{Kocevski24} also showed that $70\%-80\%$ of their photometrically-selected LRDs exhibit broad Balmer emission lines in their spectra, in agreement with previous estimates from \cite{greene23}.

Interestingly, LRDs (and JWST-discovered high-redshift AGN in general) are typically X-ray weak, with few individual detections \citep{Kocevski24}, and strict upper limits from deep X-ray stacking \citep{yue24,maiolino24,ananna24}.  Thus high-redshift AGN have X-ray emission $\sim$1 dex lower than expected from typical Type-1 AGN \citep{lambrides24,yue24}, which could indicate heavy obscuration \citep{maiolino24}, or intrinsic X-ray weakness, possibly due to mild super-Eddington accretion onto slowly spinning SMBHs \citep{pacucci24}.

\cite{kirkpatrick23} and \cite{yang23} performed AGN selection based on the MIRI dataset.  
These MIRI selections target the mid-IR dust emission from the AGN ``waste heat'', thus tending to select obscured AGN. 
About two dozen AGN have been selected, and the AGN fraction among MIRI-selected galaxies tends to be higher toward high redshifts. 
At $z\approx 3$--5, the black-hole accretion density based on MIRI  appears to be significantly higher than the expectations from previous X-ray surveys, indicating MIRI can disclose heavily obscured AGN missed by X-ray census \citep{yang23}.
However, these MIRI results have relatively large error bars due to the limited AGN sample size. 
Future wide-area MIRI surveys such as MEGA (PI: A.\ Kirkpatrick) and MEOW (PI: G.\ Leung) can overcome this disadvantage and probe obscured AGN at even higher redshifts ($z>5$).

\subsection{Galaxy Assembly}
One of the major goals of CEERS was to quantify the evolution of the rest-frame optical structure of galaxies from the epoch of reionization to cosmic noon. After the initial data were taken, we analyzed the NIRCam morphologies of a sample of 850 galaxies at $z=3-9$ that were initially identified by the CANDELS survey \citep{kartaltepe23}. We found that galaxies at these redshifts have a wide diversity of morphologies, with galaxies with disks making up $\sim60\%$ at $z\sim3$ to $\sim30\%$ at $z>6$ (see also \citealt{robertson23}). This is a larger fraction than identified by \textit{HST} at these redshifts (e.g., \citealt{kartaltepe15}) driven by the significant difference in depths (low surface brightness disks are much easier to detect with \textit{JWST}), the different rest-frame wavelengths probed, and the high resolution of \textit{JWST}, enabling the highest redshift galaxies to be resolved for the first time. Whether all of these visually identified disks are truly kinematic rotating disks is an open question, with some evidence suggesting that many might instead be intrinsically prolate (e.g., \citealt{vega24, pandya24}).

We also identified potential galaxy mergers and interactions among this sample of $z=3-9$ galaxies visually, and tested two different machine learning algorithms using these visually identified mergers as well as those from the IllustrisTNG simulation \citep{rose23} and found that these algorithms hit a ceiling of correctly classified mergers (and non-mergers) of $\sim60-70\%$ \citep{rose24}. This ceiling is likely due to the difficulty of identifying mergers via morphological signatures in systems with various merger phases, mass ratios, and gas fractions. A cosmological simulation with a larger volume (and therefore a larger number of mergers) and finer time steps would enable this to be tested further.

Stellar bars play a crucial role in the secular evolution of disk galaxies by driving gas inflows to the central region, triggering intense star formation and building bulges (e.g., \citealt{Athanassoula2003, kormendy04, jogee05, Sellwood2016}). Over the last two decades, most \textit{HST} studies explore bars out to $z \sim 1.2$. The sensitive, high-resolution near-infrared images from \textit{JWST} NIRCam allow us, for the first time, to study bars at $z >2$. EGS-23205 is one of the very first barred galaxies discovered at $z>2$ with \textit{JWST} (\citealt{guo23}). The bar structure is prominent in the \textit{JWST} NIRCam F444W image, which traces the stellar structure in the rest-frame near-infrared light at $z\gtrsim 2$ (Figure 9). The projected semi-major axis of the bar is $\sim$ 3 kpc with a projected bar ellipticity of $\sim$ 0.41 (\citealt{guo23}).  The presence of such a well-developed bar in a disk galaxy at $z\sim 2$ (with another seen later at $z \sim$ 3, \citealt{constantin23b}), when the Universe was less than 4 billion years old, is remarkable:  it suggests that bar-driven secular evolution comes into effect at least as early as $z \sim 2$  and it challenges theoretical models by showing that bars can exist in early $z \sim 2$ disk galaxies that may be significantly different from their present-day counterparts in terms of their gas fraction, turbulence, and the extent to which they are dynamically cold. 

The CEERS data have also enabled new studies of the 3D geometry of early galaxies building off of the long history of similar {\it HST} studies \citep[e.g.,][]{vanderwel14b,zhang19}. \citet{pandya24} showed that low-mass ($\log M_*/M_{\odot}=9-10$) galaxies at $z>1$ in CEERS preferentially appear elongated with typical projected axis ratios of $b/a\approx0.4$. These low-mass, high-redshift galaxies trace out a ``banana’’ in the $b/a-\log a$ diagram with an excess of low $b/a$ (edge-on) objects and a deficit of high $b/a$ (face-on) systems, particularly at larger sizes. This is naturally expected if early dwarf galaxies have prolate or triaxial 3D shapes, i.e., they must be significantly flattened along two axes unlike spheroids or circular disks. \citet{pandya24} showed that CEERS is complete to face-on circular disks over a reasonable range of sizes and magnitudes but stressed the need to scrutinize even deeper surveys. If confirmed, the predominance of elongated early galaxies can be reconciled with theory if these systems are the result of mergers happening along a preferential direction, i.e., along filaments of the cosmic web \citep{ceverino15b,tomassetti16,pozo24}. \citet{pandya19} proposed that in such a scenario, strong intrinsic alignments are expected between elongated galaxies. In a follow-up CEERS paper, \citep{pandya24b} showed that the shear from both galaxy-galaxy and weak gravitational lensing is not large enough to explain the excess of early elongated galaxies. However, they detected strong alignments in multiple NIRCam chips, modules and pointings, which they attributed to ``cosmic shear’’ from an overdensity at $z\sim0.75$ though could not rule out intrinsic alignments, PSF systematics, and other possible biases.

\subsection{Galaxy Properties} \label{gal-property}

Analyses of the early CEERS data showed that the galaxies near and into the epoch of reionization have steep UV spectral slopes (blue colors) indicating young stellar populations, low dust attenuation, and strong nebular emission at the earliest times.  \citet{whitler23b} showed that models of galaxy SEDs required relatively young ages of 30--70~Myr, implying they are dominated by young, short-lived stellar populations and rapidly rising star-formation histories. Turning back the clock on the star-formation histories, \citet{whitler23b} argued that only 3\% of the $z\sim 8-10.5$ galaxies would be UV luminous, unless they experience bursts, which foreshadowed results to come. \citet{cullen23} measured the UV spectral slopes of $z > 8$ galaxies using the CEERS NIRCam data, finding they are very blue, similar to the 0.3$Z_\odot$ starburst galaxy NGC 1705.  Their measured distribution of UV spectral slopes for $z > 8$ galaxies was similar to those at $z\sim 5$, suggesting no strong evolution over this redshift range.   \citet{endsley23c} modeled the CEERS data for $z\sim 7-8$ galaxies, showing that lower luminosity/lower mass galaxies have higher specific SFRs, accompanied by evidence for nebular emission. \citet{papovich23} used CEERS/MIRI imaging to probe rest-frame wavelengths out to $\sim$1~\micron\ for galaxies  at $4 < z < 9$.  They demonstrated the galaxies have mass-to-light ratios lower than expected, and consistent with stellar populations dominated by short-lived stars.  
 The analysis of galaxy star-formation histories from CEERS data showed that that galaxies at $z > 4$ become ``burstier'' as a function of increasing redshift and increasing stellar mass \citep{cole23}, which is interpreted as changes in the  timescales of gas accretion and the strength of feedback. Similar results have been found with other \JWST\ imaging surveys \citep{Ciesla23}.   
   
Other studies with CEERS focused on rarer, red galaxies, including populations completely undetected in \hst\ imaging.  
Largely these analyses characterized  galaxies selected with red NIRCam colors \citep{bisigello23,perezgonzalez23,barro24}. 
\citet{perezgonzalez23} selected red galaxies with  F150W $‑$ F356W $>$ 1.5 mag finding most have SEDs consistent with dusty star-forming galaxies with photometric redshifts $2 < z < 6$, though a smaller fraction are candidates for quiescent galaxies at $3 < z < 5$ or higher redshift galaxies with extremely strong emission line equivalent width.  
\citet{bisigello23} identified galaxies with F200W $-$ F444W $>$ 1.2.  
They found that while nearly three-quarters of the population appear to be lower-mass galaxies at $z < 2$, heavily extincted by dust, with $A(V) > 4$~mag, there remained a population of $z > 3$ with large dust attenuation, similar to the sample of \citealt{perezgonzalez23}. \citet{carnall23} identified quiescent galaxies at $3 < z < 5$ in the CEERS/NIRCam imaging, finding these have stellar masses of $\log M_\ast/M_\odot > 10.1$. The star-formation histories of these galaxies imply formation epochs as early as $z\approx 10$.  Moreover, \citet{carnall23} argued that the pre-\JWST\ predictions  underestimated the number density of this population by as a factor of 3--5. 

\citet{barro24} identified galaxies with F277W $‑$ F444W $>$ 1.5 mag, which pushed the selection of red galaxies to higher redshifts.    They modeled the available NIRCam, MIRI, and NIRSpec data and found that most of these galaxies have photometric redshifts $5 < z < 7$ all SEDs suggestive of heavily obscured star-formation or AGN.  \citet{barro24} point out that all the galaxies the galaxies in their sample  unresolved in the NIRCam images, which may relate them to ``Little Red Dots''.

\begin{figure*}[!t]
\centering
\epsscale{1.1}
\plotone{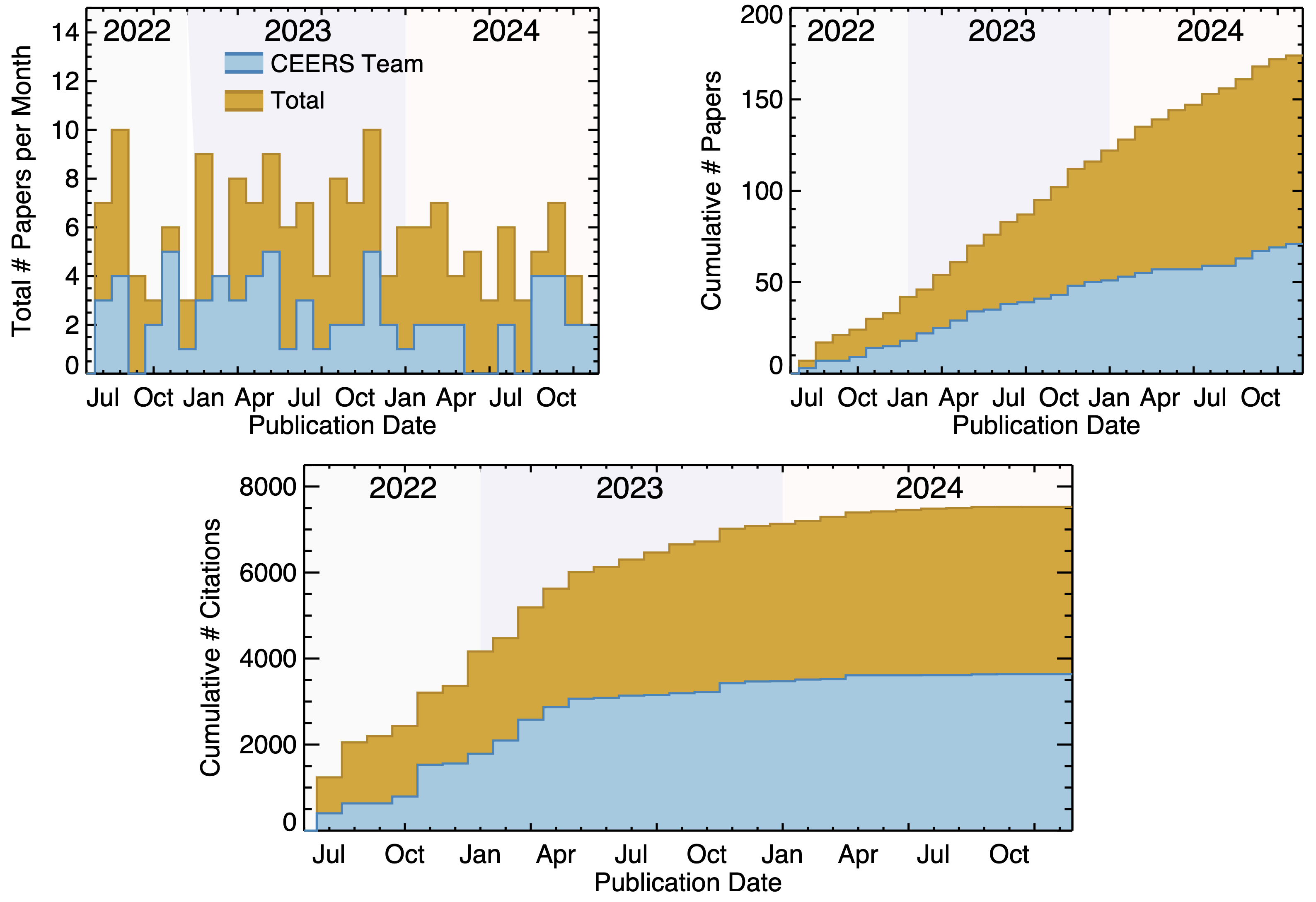}
\caption{A summary of publication statistics for papers using CEERS data, separated by papers written by CEERS team members (blue) and those written by the community (gold); the community results are stacked on top of the CEERS team results.  The upper-left panel shows papers per month, while the upper-right panel shows the cumulative paper totals.  As of December 2024, 174 papers have been written using CEERS data -- 71 by CEERS team members and 103 by the community.  The bottom panel shows that these papers have accumulated $>$7500 citations by Dec 2024, roughly half from community papers.  These statistics illustrate the powerful utility of widely available public data sets.}
\label{fig:pubstats}
\end{figure*}

CEERS spectra have also been used by many in the community to uncover the physical properties of galaxies.  \citet{sanders23} stacked galaxies by redshift from $z =$ 2–9, finding that line ratios in galaxies at $z =$ 2–6.5 are all offset from $z =$ 0 galaxies in a manner reflecting a harder ionizing spectrum at fixed metallicity (reflective of young, alpha-enhanced stellar populations [\citealt{steidel16}] and/or populations with a higher ionization parameter).  This had already been observed at $z =$ 2, but CEERS spectra allow these ionization-sensitive rest-optical lines to be observed out to higher redshift.  Interestingly, they did not find evidence for further evolution from $z =$ 2 to $z =$ 6.  In \citet{sanders24}, they followed this up by measuring metallicities based on the direct-Te method with detections of the auroral [\ion{O}{3}] 4363 \AA\ emission line for a sample of 25 galaxies (including 16 from CEERS) at $z > 2$, deriving new strong-line metallicity indicators for use at high redshift.  \citet{shapley23b} used CEERS spectra to measure the dust-sensitive Balmer decrement, finding little evolution in the dust attenuation at fixed stellar mass from $z \sim$ 2 to $z \sim$ 6 (\citealt{reddy23} also showed, using Paschen-line ratios, that Balmer-line-based dust attenuation ratios are robust).  Using strong-line-ratio metallicity diagnostics, \citet{shapley23} also explored the mass-metallicity relationship over this same redshift range.  Combined with results from \citet{nakajima23}, these data indicate little evolution in this relationship over this time interval.  Collectively, these observations show the interesting result that after significant evolution in typical ISM properties over the $\sim$10 Gyr from $z =$ 0 to $z =$ 2 (see also results from the CEERS grism data in \citealt{backhaus24}), there is significantly less evolution over the next $\sim$2 Gyr from $z \sim$ 2 to $z \sim$ 6.

The sensitivity of the CEERS photometry combined with the incredible strength the of H$\alpha$ and [\ion{O}{3}] at $z>4$ has enabled studies of this emission from these lines solely from identification in the broad-band photometry. Sources in the tail of the highest equivalent width (EW) sources are dubbed extreme emission line galaxies (EELGs).  Identification of EELGs \citep{davis24, llerena2024} reveals that many of these galaxies have rest-frame EWs $>$2500 \AA, with a handful of photometrically derived candidates at $>$4000 \AA\ which do not match predictions from models that do not include AGN or recent bursts in star formation activity. Physical properties of these EELGs reveal evidence for AGN contribution, generally high specific SFRs, strong ionizing photon production efficiency, and indications that these EELGs are candidates for LyC leakers.  

The combined sensitivity of the CEERS NIRCam and MIRI photometry has enabled the identification and mapping of regions of optically-thick dust extinction inside star-forming galaxies at redshift $\sim 3$. Exploiting the comparative degeneracy of the extinction law at wavelengths $\lambda\gtrsim0.7$ $\mu$m relative to the UV/Optical window, \citet{cheng24} used spectral population synthesis modeling to detect the presence of optically-thick dust from the excess over the best-fit SED at $\lambda >$ 1 $\mu$m when $\tau\lesssim 1$ at those wavelengths. Thanks to NIRCam's angular resolution, they could map the regions with $\tau > 1$, finding that these are often not located in the center of the galaxies, ruling out emission from potential AGN (although the AGN might contribute in those cases where the excess is centrally located). They find that the presence of optically-thick dust is often correlated with recent bursts of star formation in the SFH of the galaxies, and that $\approx 35$-50\%, and up to a factor $\approx 2$,  of stellar mass and SFR are not accounted from SED modeling in presence of optically-thick absorption.

The contribution from the thermally pulsing asymptotic giant branch (TP-AGB) stellar phase in the near-infrared (NIR) rest-frame spectra of young quiescent galaxies has been controversial for decades, as it impacts derivation of ages and masses from fitting evolutionary population synthesis models. Based on CEERS \textit{JWST}/NIRSpec observations,  \citet{lu24} report the first detection of strong cool-star signatures in the rest-frame NIR spectra of three young ($\sim$1Gyr), massive ($\sim$10$^{10}$ M\sol) quiescent galaxies at $z =$ 1--2. One of them, D36123, exhibits an exceptionally high-quality PRISM spectrum (SNR$>$187),  showing features unequivocally ascribed to TP-AGB stars (Figure~\ref{fig:specfig}). The co-existence of oxygen- and carbon-type absorption features, spectral edges and features from rare species such as Vanadium, and possibly Zirconium, reveal a strong contribution of TP-AGB stars to the NIR rest frame. Population synthesis models with significant TP-AGB contribution reproduce the observations better than those with weak TP-AGB, pointing to lower masses and younger ages. However, no existing model can fit the observed spectra well, thus suggesting that developing improved models in the future might be needed.

\subsection{Publication Analysis}

Here we analyze the impact of the CEERS program by using the NASA Astrophysics Data Server (ADS) to explore publication statistics for papers which made use of CEERS data.   We created two ADS libraries, one for papers led by CEERS team members, and one for papers led by non-CEERS members of the community.  To allow for automated identification of such papers, we used simple search criteria — if the word “CEERS” appeared in the abstract, we considered it to be highly likely to have used the CEERS data in some way.  If the CEERS PI appeared in the author list, then we considered it to be a team paper.  The results of this automated search were then (lightly) curated to remove any obvious non-relevant papers.  The analysis done here includes all papers published or posted to arXiv by December 2nd, 2024 (roughly two years to the date from when the majority of the CEERS data became available for download).

We show the results of our analysis in Figure~\ref{fig:pubstats}.  In the first panel, we show the number of papers per month, showing the total number of papers in gold, with those led by CEERS team members in blue (where the difference is papers led by community members).  In the top-right panel, we show the cumulative number of papers per month.  Both the CEERS team and total distributions show a steady rise, with a total publication rate steady at $\sim$6 papers per month.  Overall, the CEERS team has written 71 papers, while the community has led 103 papers.  In total the CEERS data have yielded 174 peer-reviewed (or submitted) publications to date; this is already a yield of $\sim$2.3 papers per hour of investing observing time (considering the final executed CEERS program time of 77.2 hours).
The final figure shows the cumulative citations.  Over the past two years, these 174 CEERS papers have yielded $>$7500 citations in total, with roughly half this number coming from the papers led by CEERS team members.

These numbers demonstrate the tremendous impact achieved by the CEERS program.  This was due to a variety of factors, including the public nature of the ERS data, the fact that (some of the) CEERS NIRCam data were amongst the first \textit{JWST} data obtained, and that CEERS provided data of a type and quality useful for a wide range of extragalactic analyses.  The continued publication of papers using this now 2.5-year-old dataset, and subsequent followup programs (e.g., RUBIES \citep{degraaff24}, CAPERS [PI Dickinson], MEGA [PI Kirkpatrick]) shows that CEERS will continue to have a tremendous legacy value.

\section{Conclusions}
In this paper we have presented the motivation behind the Cosmic Evolution Early Release Science Survey.  As one of 13 approved ERS programs, CEERS was designed to allow a variety of science investigations into galaxy evolution across cosmic time while also testing efficient coordinated-parallel operations of JWST instruments.

This 77.2 hour program generated imaging data with NIRCam and MIRI, and spectroscopic data with NIRSpec (MSA) and NIRCam (slitless).  The NIRCam imaging data reaches point-source limiting magnitudes of $\sim$29--29.5 across 1--5$\mu$m.  The MIRI imaging reaches a depth of $\sim$26th mag at $\lambda <$ 10$\mu$m, and $\sim$22-23rd mag at $\lambda \sim$ 20$\mu$m.  The NIRSpec spectroscopic data reach limiting emission line fluxes of $\sim$1--2 $\times$10$^{-18}$ erg s$^{-1}$ cm$^{-2}$ with the medium-resolution gratings, while the prism detects continuum emission to m $\sim$ 26 (as well as brighter emission lines).  The NIRCam WFSS data detect emission lines to $\sim$5--8 $\times$10$^{-18}$ erg s$^{-1}$ cm$^{-2}$.  These sensitivities are modestly deeper than pre-launch expectations, consistent with {\it JWST} post-launch documentation \citep{rigby23}.

The CEERS team has distributed high-level reduced data products from all instruments and observing modes (via our website and MAST).  As one of the first science programs executed with {\it JWST}, our reduction pipeline by necessity included several custom procedures, and we document our entire reduction process in detailed documentation and Python notebooks distributed with our data releases.

The CEERS data have enabled a wide range of science investigations, with some of the highlights discussed above including:
\begin{itemize}

\item Amongst the first discovery of numerous $z >$ 10 galaxy candidates, many now spectroscopically confirmed via CEERS and associated followup programs.

\item While the abundance of both UV-luminous galaxies and massive galaxies are not in significant tension with $\Lambda$CDM, the abundance at higher redshifts implies evolution in physics regulating star formation.

\item The discovery of accreting super-massive black holes existing in otherwise ``normal" galaxies, the first example of a high-redshift ``little red dot", and detailed investigation into AGN at moderate redshifts via dust emission detected with MIRI.

\item A demonstration that {\it JWST} can probe reionization via the detection of Ly$\alpha$ well into the epoch of reionization.

\item The discovery of the first barred spiral galaxies at $z \gtrsim$ 2.
\end{itemize}

The CEERS data have resulted in $\sim$170 papers (at publication of this paper), a majority led by non-team members, on a wide range of topics involving galaxy evolution from $z \sim$ 1--14, accumulating $>$7500 citations to date.  The significant science flowing from the CEERS program celebrates the success of the full DD-ERS program, showing the utility of publicly available data.  

We conclude by reflecting on the nature of these public ERS programs, which have been an enormous success across the board. While the urge to be the first to discover a new phenomenon is inherent in our drive to be scientists (and the CEERS team was hardly immune to this desire), we aimed to be a collaboration which supported people first.  The CEERS team used a firm but liberal publication policy to allow freedom of investigation, while attempting to protect the interests of junior researchers.  We supplemented this with a variety of open communication channels (telecons, Slack, meetings) to encourage frequent and open communication about all projects.  The outcome of this supportive environment was that a vast majority (56/71) of the CEERS-team-led papers were led by junior scientists.  Thus in addition to demonstrating the exquisite performance of {\it JWST}, ERS programs like CEERS are serving as a springboard to the careers of future leaders in astrophysics.

\begin{acknowledgements}
We acknowledge that the location where this work took place, the University of Texas at Austin, sits on indigenous land. The Tonkawa lived in central Texas and the Comanche and Apache moved through this area. We pay our respects to all the American Indian and Indigenous Peoples and communities who have been or have become a part of these lands and territories in Texas, on this piece of Turtle Island. 

We thank Alyssa Pagan (STScI) for the creation of the color image used in Figure 1, and the NIRSpec instrument team for their help with the IPS simulations.  We thank the engineers and scientists at NASA/ESA/CSA and STScI for building, launching and operating this fabulous observatory, and thank STScI for their long-standing support of the ERS program.  We thank the scientists at STScI for creating, planning and implementing the ERS program, including Ken Sembach, Janice Lee, Neill Reid, and Amaya Moro-Martin.  We also thank community members for their efforts in doing outstanding science with these data.
The authors acknowledge the Texas Advanced Computing Center (TACC, \url{http://www.tacc.utexas.edu}) at the University of Texas at Austin for providing HPC and visualization resources that have contributed to the research results reported within this paper. 
We acknowledge support from NASA through STScI ERS award JWST-ERS-1345, JWST-GO-2750, JWST-AR-2687, and JWST-AR-1721.  D.\ B.\ and M.\ H.-C.\ thank the Programme National de Cosmologie et Galaxies and CNES for their support.  RA acknowledges support from the Severo Ochoa grant CEX2021-001131-S funded by MCIN/AEI/10.13039/501100011033. 
\end{acknowledgements}

\begin{appendix}

\section{CEERS Key Papers}

In this section we list the Key Papers written by CEERS team members.  Key papers were proposed by CEERS team members prior to the launch of {\it JWST}, with papers decided on by the CEERS Executive Committee.  These papers were intended to be written with significant involvement from the full team on high-priority topics, and thus were the only papers where team publication policy prohibited internal competition on a given topic (until the Key Paper was submitted).  Nine key papers have been written, and one or two more on the CEERS spectroscopy are anticipated.

\begin{itemize}
    \item CEERS Key Paper I. An Early Look into the First 500 Myr of Galaxy Formation with JWST \citep{finkelstein23}.
    \item CEERS Key Paper II. A First Look at the Resolved Host Properties of AGN at 3 $< z <$ 5 with JWST \citep{kocevski23}.
    \item CEERS Key Paper III. The Diversity of Galaxy Structure and Morphology at $z =$ 3--9 with JWST \citep{kartaltepe23}.
    \item CEERS Key Paper IV. A Triality in the Nature of HST-dark Galaxies \citep{perezgonzalez23}.
    \item CEERS Key Paper V. Galaxies at 4 $< z <$ 9 Are Bluer than They Appear–Characterizing Galaxy Stellar Populations from Rest-frame $\sim$1 $\mu$m Imaging \citep{papovich23}.
    \item CEERS Key Paper VI. JWST/MIRI Uncovers a Large Population of Obscured AGN at High Redshifts \citep{yang23}.
    \item CEERS Key Paper VII. JWST/MIRI Reveals a Faint Population of Galaxies at Cosmic Noon Unseen by Spitzer \citep{kirkpatrick23}.
    \item CEERS Key Paper VIII. Emission-line Ratios from NIRSpec and NIRCam Wide-Field Slitless Spectroscopy at $z >$ 2 \citep{backhaus24}.
    \item CEERS Key Paper IX. Identifying Galaxy Mergers in CEERS NIRCam Images Using Random Forests and Convolutional Neural Networks \citep{rose24}.
\end{itemize}

\section{Additional CEERS Layouts}
\begin{figure*}[!t]
\centering
\epsscale{0.57}
\plotone{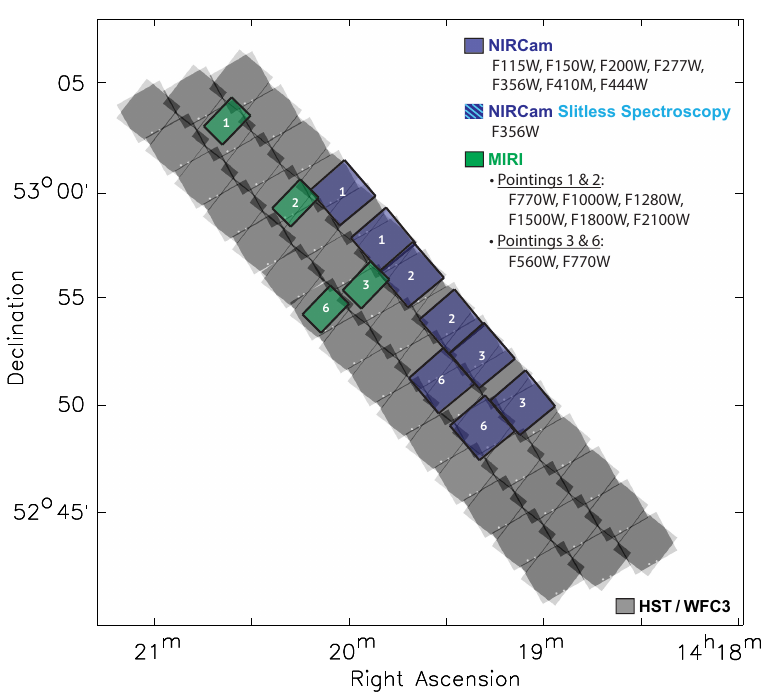}
\plotone{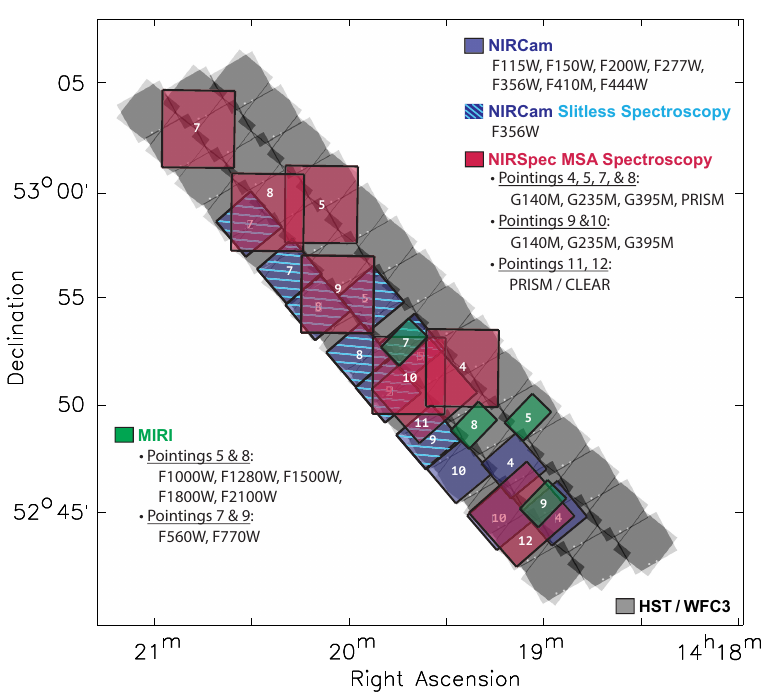}
\caption{Launch delays forced us to split the program into two epochs.  These figures show the pointings executed in June (left) and December (right; also including the rescheduled NIRSpec observations in February 2023).}
\label{fig:layout_junedec}
\end{figure*}

\begin{figure*}[!t]
\centering
\epsscale{0.57}
\plotone{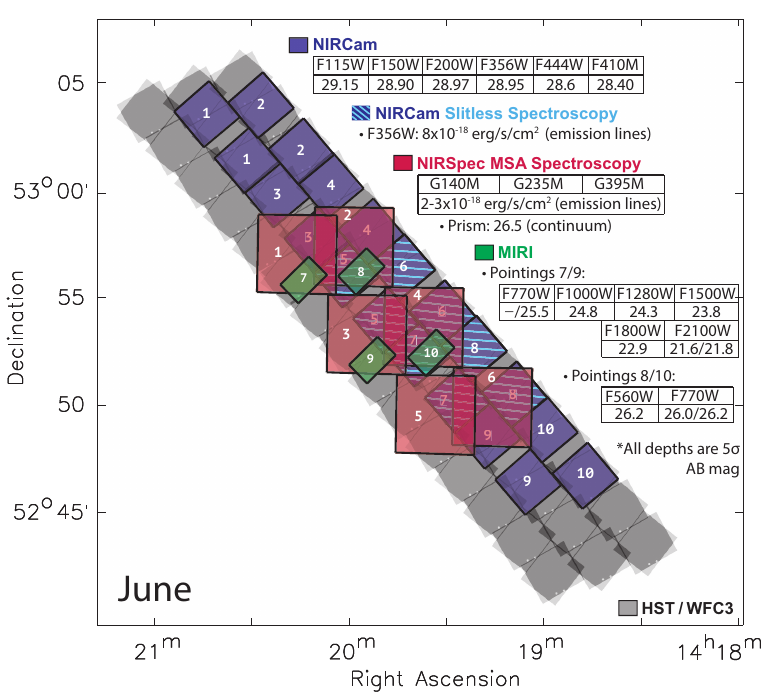}
\plotone{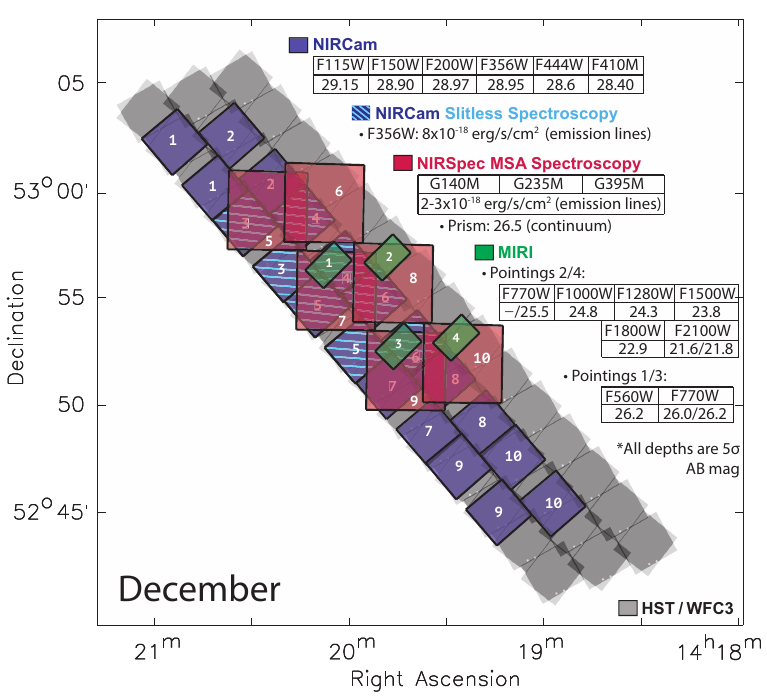}
\caption{Alternate potential single-epoch layouts for CEERS.  The left image shows that originally proposed in 2017, for observations completely within the June window.  Previous launch delays had us consider executing the program all in the December window, which we show a plan for in the right panel.  In the end CEERS was split into two epochs, as shown in Figure~\ref{fig:layout_junedec}.}
\label{fig:layout_alternate}
\end{figure*}

In this section, we include additional layout figures.  Figure~\ref{fig:layout_junedec} shows the two executed epochs separately, showing our June NIRCam+MIRI on the left, and December NIRSpec+NIRCam, and NIRCam WFSS+MIRI on the right.  Figure~\ref{fig:layout_alternate} shows the original single-epoch layouts for CEERS should scheduling have allowed the entire program to be executed in one window, as originally planned.

\section{v1.0 Data Reduction}
\subsection{NIRCam Imaging v1.0 Data Reduction}
\citet{bagley23b} presented the initial CEERS/NIRCam imaging reduction methods
and quality analysis. We describe here the updates we have implemented to our
procedures for the v1.0 release, and refer the reader to \citet{bagley23b} for
all other details. The v1.0 reduction uses an updated Calibration Pipeline
(v1.13.4) as well an updated CRDS context (jwst\_1195.pmap) that includes many
new in-flight reference files (new long wavelength darks, flats, and
distortion, readnoise and superbias reference files). We use all pipeline
default parameter values for the Stage 1 detector-level corrections and note
that snowball flagging within the Jump step of the pipeline was still turned
off by default for pmap 1195. Instead, we use our custom routine as described
in \citet{bagley23b}, which has a similar behavior as the pipeline routine,
though is optimized for larger snowballs. As a result, some of the smallest
snowballs are not flagged and corrected in our reduction.

Following Stage 1, we perform improved wisp and $1/f$ noise subtractions.
We use the wisp templates created by the JADES team
\citep{rieke23b,tacchella23b}, as these templates include almost all
of the wisp features present in CEERS images. However, we zero out the
templates off-wisp so we do not introduce extra noise to the images during
wisp subtraction. We scale and subtract the modified wisp templates by a
coefficient that minimizes the variance of the image as described in
\citet{bagley23b}.
For the $1/f$ correction, we now use a three step procedure. First, we mask
source flux using an aggressive mask and measure and subtract a 2D background
model, which captures the diffuse residuals of wisps and unmasked source flux
around bright, extended sources. Next, we measure a sigma-clipped median
value for each row of pixels, followed by each column. For this measurement,
we use a slightly less aggressive source mask -- created from only the mosaic
for the given filter -- to ensure there are sufficient unmasked pixels for a
robust measure of the striping pattern.
Finally, we measured a sigma-clipped median value for each ``amp-row,'' or
a row of pixels within a single amplifier. This second pass is aimed at
removing any remaining amp-dependent noise patterns, and the amplitude of
this correction is much lower. If too many pixels in a given amp-row were
masked, we do not apply a second correction for that amp-row.
The threshold for the number of masked amp-row pixels can vary image
to image from $\sim25-75$\%, as images with bright sources require a
different threshold than those with mostly fainter sources.

We have also redone the astrometry for all images using the JWST HST
Alignment Tool \citep[\texttt{JHAT};][]{jhat}. The absolute reference catalog is
generated from a 30mas F814W mosaic that has been tied to Gaia-EDRS
cite cite cite. Rather than aligning all filters individually to F160W as
with earlier CEERS reductions, for v1.0 we first align F277W to F814W,
choosing this ACS filter for its higher resolution. We then create a full
mosaic and catalog in F277W, and use that as the reference for all other
filters. Both the F814W and F277W catalogs are generated with Source
Extractor using windowed centroid coordinates. We also remove point
sources and spurious sources detected around diffraction spikes and
image edges from these reference catalogs. After images are aligned
using \texttt{JHAT}, we identify portions of the mosaic with median residual
offsets in right ascension and/or declination of $>$3 mas and apply
the median shift to the corresponding set of images to correct for
this residual offset. The RMS of the alignment is $\sim$5--8 mas
(NIRCam-to-NIRCam) and $\sim$8-11 mas (ACS-to-NIRCam).

Our last major change from the v0.5 and v0.6 NIRCam imaging reductions
was an improved handling of outliers, bad pixels, and persistence.
The outlier detection step of the Stage 3 pipeline is run with two
changes to the default values. We exclude pixels flagged as
\texttt{DO\_NOT\_USE} and \texttt{UNRELIABLE\_SLOPE} when building the
weight map. We also lower the fraction of maximum weight to use for
valid data (\texttt{maskpt}) from 0.7 to 0.5, and exclude one high-valued
pixel in each stack when creating the median image (\texttt{nhigh=1}).
We use the \texttt{Snowblind} routine \citep{snowblind} to identify cases of persistence.
Pixels flagged as saturated in one exposure are also flagged in each
subsequent exposure taken within 2500 seconds. We run this routine
on an association of all images in a visit in order to catch persistence
for a given detector across all filters.
Next, in each exposure, we flag the plus signs that form around RC pixels
(pixels with a large non-linear dark signal). These are identified as pixels
immediately adjacent to bad pixels in the DQ array that are also $>3\sigma$
above the sigma-clipped median of the science image.
Finally, we create additional custom bad pixel masks for each CEERS
NIRCam pointing, consisting of pixels that were flagged as jumps in 100\%
of images and/or as outliers in 90\% of images in the pointing for a
given detector. In practice, these custom maps only catch a handful
of additional bad pixels per pointing/detector.
The remainder of our v1.0 reduction follows the same steps as described
in \citet{bagley23b}. Specifically, we remove a background pedestal from each
image; rescale the variance maps to include the RMS of the sky fluctuations
measured in the science images; resample all filters onto a 30 mas output
grid; and perform a robust 2D background subtraction using a tiered source
mask (tuned to detect both extended and compact sources) that is merged
from all available NIRCam and \hst\ filters.

\subsection{NIRCam WFSS v1.0 Data Reduction}
The processing of the CEERS WFSS data relied on the Simulation Based Extraction method \citep[SBE]{pirzkal17} followed closely what is described in \citet{pirzkal24}. While the later described the process used to process NIRISS WFSS observations using, it also describes the steps we followed to extract the CEERS NIRCam WFSS  data. To briefly summarize, we relied on the v1.0 official official imaging CEERS mosaics, catalog, and object segmentation map to drive the extraction process. Individual F356W imaging mosaics of the CEERS NIRcam WFSS were first created. As the latter were astrometrically consistent with the WFSS observations taken during the same visit, a set of affine transformations (shift and rotation) were derived between the CEERS v1.0 mosaics and these shallower F356W mosaics. The pixel based flux information over a wide range of wavelength from the v1.0 mosaics could then be used to create accurate simulations of each individual WFSS observation. These simulations, which therefore included the pixel based spectral energy distribution of each pixel within the segmentation footprint of each source were then used to quantitatively estimate the spectral contamination caused by overlapping spectra. These simulations were also used to mask out regions of our data that contained dispersed spectra so that these masked data could be combined to create an estimate of the F356W dispersed background at the time and specific 6 pointings of the CEERS WFSS observations. As we described in \citet{pirzkal24} we used an additional average, cross dispersion fit of the background residuals to further improve the background subtraction. We relied on the latest official STScI NIRCam WFSS calibration products as of  August 2024. These calibrations show a good match to the data in terms of trace location (within $\approx 0.1$\ pixel) and wavelength calibration (within $\approx 2\AA$. The agreement between the R and C grisms are also excellent and within a small fraction of a pixel ($\approx 0.2$\ pixel), allowing us to use both the R and C grisms to reliably determine the observed wavelength of emission lines. The final 1D spectra were created using an optimal SBE extraction which is described in \cite{pirzkal18,pirzkal24}.

\subsection{MIRI v1.0 Data Reduction}
\cite{yang23b} presented an initial release of the CEERS/MIRI data, which was reduced based on pipeline v1.10.2 with CRDS context jwst\_1077.pmap.
We have now re-reduced the raw MIRI data using pipeline v1.12.0 with CRDS context jwst\_1130.pmap, which is included in the CEERS v1.0 data release. 
The \cite{yang23b} release includes photometric catalogs produced by Source Extractor (for the blue pointings) and TPHOT (for the red paintings) without aperture corrections. 
The v1.0 release includes aperture corrections applied to both catalogs. 
For the Source Extractor catalog, the correction factor is derived as 1/EEF($r=\sqrt{A\times B}$). 
The denominator is the PSF encircled energy fraction curve at $r=\sqrt{A\times B}$, where A and B are the semi-major and semi-minor axes of the Kron aperture used in the sextractor photometry for the source.  
For TPHOT, the correction factor is calculated as 1/EEF($r=1.5''$). The $1.5''$ radius is empirical, based on our simulated MIRI data as presented in \citet{yang21}.
In the calculations above, the EEF value for each band is derived from a composite PSF:
within $r=16.5''$ the empirical PSF from \cite{libralato24} is available and we adopt it; 
beyond $r=16.5''$ we use the WebbPSF-generated model PSF.

\end{appendix}

\bibliographystyle{aasjournal}
%\bibliography{stevenf}

% also needed for moving affiliations to the end of doc
\allauthors 

\end{document}